\documentclass[10pt,journal,compsoc]{IEEEtran}
\IEEEoverridecommandlockouts

\newlength{\DepthReference}
\newlength{\HeightReference}
\usepackage{enumitem}
\usepackage{cite}
\usepackage{amsmath,amssymb,amsfonts}
\usepackage{algorithmic}
\usepackage{graphicx}
\usepackage{textcomp}
\usepackage[dvipsnames]{xcolor}
\usepackage{siunitx}
\usepackage{xspace}
\usepackage{graphbox}
\usepackage{mathtools}
\usepackage[most]{tcolorbox}
\usepackage{fancybox}
\usepackage{makecell}
\usepackage{multirow}
\usepackage{booktabs}
\usepackage{tikz}
\usepackage[hyphens]{url}
\usepackage{hyperref}
\hypersetup{hidelinks}
\hypersetup{breaklinks=true}
\usepackage{cleveref}
\usepackage[switch]{lineno}

\usepackage[dvipsnames]{xcolor}
\usepackage{adjustbox}

\usepackage{calc}
\usepackage{algorithm}
\usepackage{array}
\usepackage[caption=false,font=footnotesize]{subfig}
\usepackage{stfloats}
\usepackage{url}
\usepackage{verbatim}
\usepackage{graphicx,xcolor}

\usepackage{tabularx} 
\usepackage{fontawesome}
\usepackage{wasysym}
\usepackage[table]{xcolor}

\usepackage{pgfplots}
\pgfplotsset{compat=1.18}
\usepgfplotslibrary{groupplots}
\usetikzlibrary{patterns}
\definecolor{mygray}{gray}{0.86}

\usepackage{xspace}

\newcommand{\bpmn}{BPMN\xspace}
\newcommand{\sm}{SM\xspace}
\newcommand{\sms}{SMs\xspace}
\newcommand{\bt}{BT\xspace}
\newcommand{\bts}{BTs\xspace}
\newcommand{\htn}{HTN\xspace}
\newcommand{\htns}{HTNs\xspace}

\newcommand\exLink[2]{[\textit{#1}~\href{#2}{\faExternalLink}]}

\newcommand{\full}{\CIRCLE}
\newcommand{\half}{\LEFTcircle}
\newcommand{\emptycirc}{\Circle}

\usepackage[most]{tcolorbox}

\newtcolorbox{observation}{%
  enhanced,
  colback=white,
  colframe=white,
  borderline west={2pt}{0pt}{black}, 
  left=3mm, right=0mm,
  top=1mm, bottom=1mm,
  boxsep=0mm,            
}

\newcommand{\findingcolor}{MidnightBlue}

\usepackage{etoolbox} 
\newcounter{obs}
\makeatletter 
\@addtoreset{obs}{section}
\makeatother

\renewcommand{\theobs}{%
  \number\numexpr\value{section}-3\relax.\arabic{obs}%
}
\newtcolorbox{obs}[2][]{%
  enhanced,
  colback=white, colframe=white,
  borderline west={2pt}{0pt}{\findingcolor},
  left=3mm, right=0mm,
  top=1mm, bottom=1mm,
  boxsep=0mm,
  before upper={\refstepcounter{obs}%
    \textbf{\color{\findingcolor}\faExclamationCircle~\textit{Observation~\theobs}~--~#2:}\par\smallskip},
  #1
}

\definecolor{agreeteal}{RGB}{11, 166, 165}
\definecolor{darkteal}{RGB}{15, 125, 124}

\definecolor{disagreeorange}{RGB}{230, 117, 27}
\definecolor{darkorange}{RGB}{209, 101, 18}

\newcommand{\CS}{\shortstack{Control\\structures}}
\newcommand{\MC}{\shortstack{Modeling\\concepts}}

\newcommand{\LikertChart}[8]{%
  \begin{figure}[t]
    \centering
    \begin{tikzpicture}
      \begin{axis}[
        xbar stacked,
        xmin=0, xmax=#1,
        width=\linewidth,
        height=3.0cm,
        bar width=10pt,
        enlarge y limits=0.35,
        xlabel={Number of responses},
        symbolic y coords={\CS,\MC},
        ytick=data,
        ytick style={draw=none},
        y dir=reverse,
        tick label style={font=\footnotesize},
        label style={font=\footnotesize},
        legend style={
            font=\footnotesize,
            at={(0.5,1.15)},
            anchor=south,
            legend columns=3
        },
        xmajorgrids,
        grid style={dotted},
        xtick distance=5
      ]
      \addplot+[
        draw=darkorange,
        fill=darkorange,
        postaction={pattern=north east lines, pattern color=white}
            ]
        coordinates {#2};

      \addplot+[fill=black!25,pattern=crosshatch dots,draw=disagreeorange,pattern color=disagreeorange]
        coordinates {#3};

      \addplot+[fill=white,draw=gray]
        coordinates {#4};

      \addplot+[fill=black!50,pattern=crosshatch,draw=agreeteal,pattern color=agreeteal]
        coordinates {#5};

      \addplot+[fill=agreeteal,draw=agreeteal,pattern color=agreeteal]
        coordinates {#6};

      \legend{Strongly disagree,Disagree,Neutral,Agree,Strongly agree}
      \end{axis}
    \end{tikzpicture}
    \caption{#7 (#1 respondents).}
    \label{#8}
  \end{figure}
}

\newcommand{\RB}{\shortstack{Reactive\\behavior}}
\newcommand{\DM}{\shortstack{Decision\\making}}
\newcommand{\TDB}{\shortstack{Time-dep.\\behavior}}
\newcommand{\TS}{\shortstack{Task\\status}}
\newcommand{\RRI}{\shortstack{Robot-robot\\interaction}}
\newcommand{\HRI}{\shortstack{Human-robot\\interaction}}
\newcommand{\RESI}{\shortstack{Robot-ext. sys.\\interaction}}
\newcommand{\SSTR}{\shortstack{State saving and\\task resuming}}
\newcommand{\EW}{\shortstack{Explicit\\waiting}}

\newcommand{\LikertChartVQTwoOne}[8]{%
  \begin{figure}[t]
    \centering
    \begin{tikzpicture}
      \begin{axis}[
        xbar stacked,
        xmin=0, xmax=#1,
        width=0.92\linewidth,
        height=10cm,
        bar width=10pt,
        enlarge y limits=0.05,
        xlabel={Number of responses},
        symbolic y coords={\RB,\DM,\TDB,\TS,\RRI,\HRI,\RESI,\SSTR,\EW},
        ytick=data,
        ytick style={draw=none},
        y dir=reverse,
        tick label style={font=\footnotesize},
        label style={font=\footnotesize},
        legend style={
            font=\footnotesize,
            at={(0.5,1.15)},
            anchor=south,
            legend columns=3
        },
        xmajorgrids,
        grid style={dotted},
        xtick distance=5
      ]

      \addplot+[draw=darkorange,
        fill=darkorange,
        postaction={pattern=north east lines, pattern color=white}]
        coordinates {#2};

      \addplot+[fill=black!25,pattern=crosshatch dots,draw=disagreeorange,pattern color=disagreeorange]
        coordinates {#3};

      \addplot+[fill=white,draw=gray]
        coordinates {#4};

      \addplot+[fill=black!50,pattern=crosshatch,draw=agreeteal,pattern color=agreeteal]
        coordinates {#5};

      \addplot+[fill=agreeteal,draw=agreeteal,pattern color=agreeteal]
        coordinates {#6};

      \legend{Strongly disagree,Disagree,Neutral,Agree,Strongly agree}

      \end{axis}
    \end{tikzpicture}
    \caption{{#7.}}
    \label{#8}
  \end{figure}
}

\newcommand{\LikertChartVQTwoTwo}[8]{%
  \begin{figure}[t]
    \centering
    \begin{tikzpicture}
      \begin{axis}[
        xbar stacked,
        xmin=0, xmax=#1,
        width=\linewidth,
        height=4.5cm,
        bar width=10pt,
        enlarge y limits=0.25,
        xlabel={Number of responses},
        symbolic y coords={\bt,\sm,\htn,\bpmn},
        ytick=data,
        ytick style={draw=none},
        y dir=reverse,
        tick label style={font=\footnotesize},
        label style={font=\footnotesize},
        legend style={
            font=\footnotesize,
            at={(0.5,1.15)},
            anchor=south,
            legend columns=3
        },
        xmajorgrids,
        grid style={dotted},
        xtick distance=5
      ]

      \addplot+[draw=darkorange,
        fill=darkorange,
        postaction={pattern=north east lines, pattern color=white}]
        coordinates {#2};

      \addplot+[fill=black!25,pattern=crosshatch dots,draw=disagreeorange,pattern color=disagreeorange]
        coordinates {#3};

      \addplot+[fill=white,draw=gray]
        coordinates {#4};

      \addplot+[fill=black!50,pattern=crosshatch,draw=agreeteal,pattern color=agreeteal]
        coordinates {#5};

      \addplot+[fill=agreeteal,draw=agreeteal,pattern color=agreeteal]
        coordinates {#6};

      \legend{Strongly disagree,Disagree,Neutral,Agree,Strongly agree}

      \end{axis}
    \end{tikzpicture}
    \caption{{#7.}}
    \label{#8}
  \end{figure}
}

\begin{document}

\title{Formalisms for Robotic Mission Specification and Execution: A Comparative Analysis}

\author{
Gianluca Filippone\thanks{The first two authors contributed equally to this paper.}, 
Sara Pettinari\thanks{}, Patrizio~Pelliccione 
\thanks{
G.~Filippone, S.~Pettinari, and P.~Pelliccione are with Gran Sasso Science Institute (GSSI), L'Aquila, Italy - e-mail: \{gianluca.filippone, sara.pettinari, patrizio.pelliccione\}@gssi.it
}
}



\makeatletter
\def\ps@IEEEtitlepagestyle{
        \def\@oddfoot{\mycopyrightnotice}
        \def\@evenfoot{}
}
\def\mycopyrightnotice{
        {\footnotesize
                \begin{minipage}{\textwidth}
                        \centering
                        \textcopyright~{\it ``This work has been submitted to the IEEE for possible publication. Copyright may be transferred without notice, after which this version may no longer be accessible.''}
                \end{minipage}
        }
}

\maketitle

\IEEEpubidadjcol

\begin{abstract}
Robots are increasingly deployed across diverse domains and designed for multi-purpose operation. 
As robotic systems grow in complexity and operate in dynamic environments, the need for structured, expressive, and scalable mission-specification approaches becomes critical, with mission specifications often defined in the field by domain experts rather than robotics specialists.
However, there is no standard or widely accepted formalism for specifying missions in single- or multi-robot systems. A variety of formalisms, such as Behavior Trees, State Machines, Hierarchical Task Networks, and Business Process Model and Notation, have been adopted in robotics to varying degrees, each providing different levels of abstraction, expressiveness, and support for integration with human workflows and external devices.

This paper presents a systematic analysis of these four formalisms with respect to their suitability for robot mission specification. 
Our study focuses on mission-level descriptions rather than robot software development. We analyze their underlying control structures and mission concepts, evaluate their expressiveness and limitations in modeling real-world missions, and assess the extent of available tool support.
By comparing the formalisms and validating our findings with experts, we provide insights into their applicability, strengths, and shortcomings in robotic system modeling. The results aim to support practitioners and researchers in selecting appropriate modeling approaches for designing robust and adaptable robot and multi-robot missions.
\end{abstract}


\section{Introduction}\label{sec:introduction}

Robots are becoming pervasive across a wide range of domains, including industrial automation, logistics, healthcare, hospitality, and agriculture~\cite{2018Fischer,10.1016/j.robot.2021.103866}. At the same time, robots are increasingly multi-purpose, capable of performing diverse tasks rather than being tailored to a single function~\cite{garcia2023software}. As a result, missions must often be specified and adapted directly in the field by domain experts, who are responsible for defining robot behavior despite not necessarily having expertise in robotics, programming languages, or computer science~\cite{dragule2025effects,Promise,TSEPatterns2021,TSEPatterns2023}.

As robotic systems are increasingly deployed in dynamic, real-world settings, several efforts have sought to specify missions using structured, expressive, and scalable formalisms~\cite{dragule2025effects,thorsten2023behavior,Promise}. Effective mission specification for single- and multi-robot systems demands approaches that balance clarity, modularity, and ease of use with the ability to adapt during execution. Despite sustained research and industrial interest, however, no standard or widely accepted formalism has emerged that adequately addresses these requirements across diverse application domains.
 
Instead, the state of the art is fragmented across multiple formalisms, including Behavior Trees (\bt)~\cite{colledanchise2021on,colledanchise2016advantages,colledanchise2018behavior,thorsten2023behavior}, State Machines (\sm)\footnote{For presentation purposes, we use the term State Machine (SM) to encompass both Finite State Machines (FSMs) and Hierarchical Finite State Machines (HFSMs).}~\cite{flexbezutell2022ros,schillinger2016flexbe,thorsten2023behavior},  Hierarchical Task Networks (\htn)~\cite{lesire2016distributed}, and Business Process Model and Notation (\bpmn)~\cite{ReyCCMC19,de2020event,whitaker2024mission,FaMe}. While \bt and \sm are widely adopted in robotics due to their relative simplicity and execution efficiency, they provide limited support for integrating human-driven tasks and external workflows. Conversely, \htn and \bpmn offer richer abstractions for coordination and integration but introduce additional modeling complexity and have yet to achieve broad adoption in robotics~\cite{robomax}.

This fragmentation is mirrored in practice. High-profile projects, such as NASA’s Europa Lander mission\footnote{\url{https://ai.jpl.nasa.gov/public/projects/europa-lander/}}, have experimented with both \htn-based planning and \bpmn-based workflow modeling to coordinate robotic activities~\cite{de2020event,nasaBasichRCZ22,nasaWangRBC22}, underscoring the absence of a dominant solution. In industrial contexts, vendors largely rely on proprietary graphical or block-based languages (e.g., Dobot, KUKA, Universal Robots), which further limits portability and reuse. Although some companies, such as PAL Robotics and Bosch, have begun exploring standard formalisms like \sm~\cite{flexbezutell2022ros}, \bt, and \htn\footnote{\url{https://docs.pal-robotics.com/ari/sdk/23.12/development/intro-development.html}}, the lack of consensus and systematic comparison continues to hinder informed selection and adoption.

Each of these formalisms offers different levels of abstraction, expressiveness, and control, which in turn shape how robotic systems are designed, verified, and executed. The choice of formalism is therefore not neutral, but depends on factors such as mission complexity, required adaptability, and execution constraints.
Prior work has begun to examine these trade-offs. For instance, \cite{thorsten2023behavior} analyzes key language concepts in Behavior Trees and contrasts them with State Machines, which remain the de facto standard for behavior modeling in robotics. Complementarily, \cite{dragule2025effects} reports a controlled experiment evaluating the effectiveness and efficiency of \bt and \sm when used by end users to specify robot missions. While these studies provide valuable insights into individual formalisms, they do not offer a comprehensive comparison across the broader space of mission-specification approaches or address their suitability for complex, real-world robotic missions.

In this work, we analyze Behavior Trees (\bt) and State Machines (\sm) from a practical perspective, focusing on their control structures and mission abstractions, as well as their distinctive characteristics, limitations, and tool support. In addition to \bt and \sm, we also consider Hierarchical Task Networks (\htn) and Business Process Model and Notation (\bpmn). For clarity, we collectively refer to \bt, \sm, \htn, and \bpmn as ``the formalisms''. By systematically comparing these approaches, we aim to provide practitioners and researchers with concrete insights into their applicability for robotic system modeling.
Our analysis assumes that robots and their underlying software components are already implemented. We therefore focus exclusively on mission description, understood as a natural-language or domain-specific specification of the activities a robot must perform~\cite{TSEPatterns2021}. To ensure conceptual clarity, we adopt the terminology introduced in RobMoSys~\cite{robmosysD21} and used in subsequent studies~\cite{garcia2023software,thorsten2023behavior}. Specifically, we refer to a {\em skill} as a programmed action executable by a robot and typically implemented as a software component; a {\em task} refers to a simple, coordinated behavior composed of multiple skills; and a {\em mission} represents a coordinated sequence of tasks that enables the robot to achieve its overall objective.

To systematically compare the formalisms and address the lack of consolidated guidance for mission specification, we structure our analysis around the following research questions:

\begin{itemize}
    \item \textit{RQ1: To what extent do the formalisms support the representation of control structures and mission concepts?}
    \\
    \emph{Rationale:} Since each formalism relies on different execution and control abstractions, understanding how they represent core mission constructs (e.g., sequencing, branching, concurrency, and coordination) is essential to assess their expressive power and suitability for mission-level modeling.

    \item \textit{RQ2: What are the peculiarities and limitations of modeling missions with the formalisms?}  
    \\
    \emph{Rationale:} While a formalism may be expressive in principle, its practical applicability depends on how naturally and effectively it supports realistic mission scenarios. This question investigates modeling trade-offs, abstraction levels, and limitations that emerge when specifying complex single- and multi-robot missions.

    \item \textit{RQ3: Which publicly available tools support the formalisms, and to what extent?}  
    \\
    \emph{Rationale:} Tool support is a key factor for real-world adoption. By examining available modeling, execution, and verification tools, this question evaluates the maturity and practicality of each formalism beyond its theoretical foundations.
\end{itemize}

Together, these research questions examine how the formalisms model mission control structures (\textit{RQ1}), reveal their practical strengths and limitations when specifying realistic robotic missions (\textit{RQ2}), and evaluate the maturity and effectiveness of available tool support for mission specification and execution (\textit{RQ3}).

We validated our findings through an expert questionnaire survey conducted according to established guidelines~\cite{acmsigsoft}, using purposive sampling~\cite{etikan2016comparison} to recruit authors of the reference works underlying our analysis. Participants self-assessed their expertise and evaluated only the formalisms they knew, rating our results on \textit{completeness}, \textit{correctness}, and \textit{alignment} via Likert-type items complemented by mandatory justifications for neutral-or-lower ratings; we then analyzed responses a posteriori and followed up with selected experts to clarify and deepen critical feedback.

{\bf Paper outline}: Section~\ref{sec:background} provides an overview of the four formalisms analyzed in this paper. Section~\ref{sec:methodology} describes the research method we defined to perform the study, together with the analysis corpus used to compare the considered formalisms. Section~\ref{sec:concepts} answers RQ1 by analyzing the formalisms in terms of control structures and mission concepts. Section~\ref{sec:expressivity} answers RQ2 by describing peculiarities and limitations of modeling missions with the formalisms. Section~\ref{sec:tools} answers RQ3 by discussing available tools supporting the formalisms. We validated the findings with experts on the formalisms. The validation of each RQ is reported in the respective section. 
Section~\ref{sec:threats} presents the threats to validity and the mitigation strategies adopted.
Section~\ref{sec:discussion} discusses the findings of the study. Section~\ref{sec:related} discusses the related works. The paper concludes with final remarks and future works in Section~\ref{sec:conclusion}.

\section{The Formalisms}\label{sec:background}

This section provides an overview of the four formalisms analyzed in this paper by providing a lightweight description of their components and semantics and briefly discussing their origins.

\subsection{Behavior Trees}

\bts were originally developed to serve the videogame industry as an approach to design the artificial intelligence of non-player characters (NPCs), as an alternate way to \sms~\cite{colledanchise2018behavior,iovino2022survey}.
In the last years, \bts gained popularity in the robotics industry and research as a modular and flexible approach to describe robots' behaviors by structuring decision-making logic hierarchically, with states represented as leaves in a tree~\cite{thorsten2023behavior}. The work in~\cite{GUGLIERMO2024104714} provides a comprehensive overview of BT functional and non-functional properties that are relevant for the robotic community, how they relate to each other and the metrics to measure BTs. 

\begin{figure}[htb]
    \centering
    \includegraphics[width=0.8\linewidth]{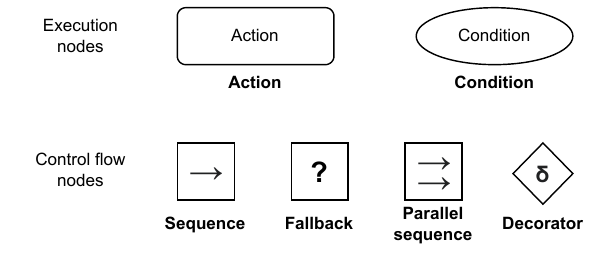}
    \caption{Core \bt\ elements.}
    \label{fig:bt_elements}
\end{figure}

A \bt{} is a directed rooted tree whose internal nodes are called \textit{control flow nodes} and leaf nodes are called \textit{execution nodes}. The execution of the tree is performed through \textit{ticks}. They are periodic signals that are sent from the root and propagate through its children. When a node receives a tick, it executes its behavior (which can be a flow control task or the execution of a robotic skill) and immediately returns to its parent node one a status: \textit{Success}, if the execution completed successfully, \textit{Running}, if the execution is currently in progress, \textit{Failure}, otherwise. In their most classical formulation~\cite{colledanchise2018behavior}, the core elements, i.e., nodes, of a \bt{} consist of two types of execution nodes (\textit{Action} and \textit{Condition} nodes) and four types of control flow nodes (\textit{Sequence}, \textit{Fallback}, \textit{Parallel}, and \textit{Decorators}), as shown in Figure~\ref{fig:bt_elements}.

\textit{Action nodes} execute specific commands. In robotics, they typically map to skills, which are reusable, parameterized behaviors such as navigating to a location, grasping an object, or manipulating a tool. When ticked, action nodes perform their associated skill and return \textit{Success}, \textit{Failure}, or \textit{Running}, as explained above.
\textit{Condition nodes} evaluate boolean expressions related to the system state, e.g., checking whether an object is detected or if a robot has reached its destination. They return \textit{Success} if the condition holds and \textit{Failure} otherwise. They never return \textit{Running} as they do not represent actions that are executed.

\textit{Control flow nodes} manage the tick propagation through the tree according to their specific semantics.
\textit{Sequence} nodes tick the children in order, returning \textit{Success}, if and only if, all its children return \textit{Success}. If a child returns \textit{Failure} or \textit{Running}, the next children are not ticked, and the node returns \textit{Failure} or \textit{Running}, accordingly.
\textit{Fallback} nodes tick the children in order as in the previous case, but return \textit{Failure}, if and only if, all its children return \textit{Failure}. Similarly, if a child returns \textit{Success} or \textit{Running}, the next children are not ticked, and the node returns \textit{Success} or \textit{Running}, accordingly.
The \textit{Parallel} node ticks all its children (possibly) simultaneously and returns \textit{Success} or \textit{Failure}, if at least a certain number of children returned \textit{Success} or \textit{Failure}, respectively.
The \textit{Decorator} node is a custom control-flow node that features only one child and whose behavior is user-defined, via a so-called policy. Typical examples of decorator nodes are the \textit{Inverter} node, which alters the child's return status, and the \textit{Repeater}, which forces repeated executions.

\begin{figure}[!b]
    \centering
    \includegraphics[width=0.75\linewidth]{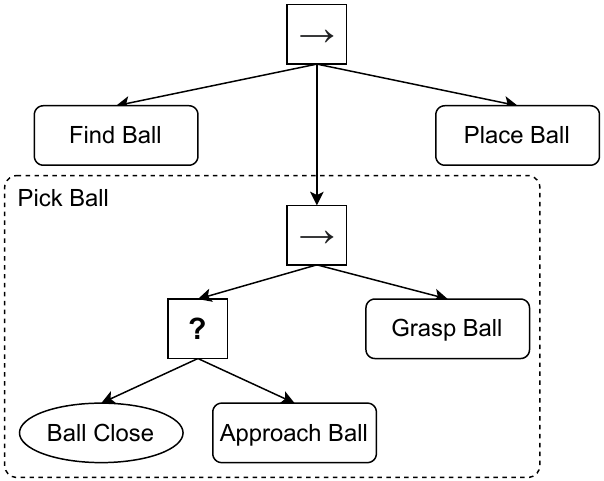}
    \caption{Example of mission expressed as a \bt.}
    \label{fig:bt_example}
\end{figure}

Figure~\ref{fig:bt_example} shows an exemplar mission expressed as a \bt{}, adapted from~\cite{colledanchise2018behavior}. Its execution is commanded by ticks that are sent to the root sequence node with a certain frequency. This node propagates the tick to the \textit{Find Ball} action node. When it returns \textit{Success}, it is propagated to the \textit{Pick Ball} subtree (see the dashed-bordered box) and, then, to the \textit{Fallback} node. From there, the \textit{Ball Close} condition is checked. If the ball is not close to the robot, the condition node returns \textit{Failure}, and the tick is propagated to the \textit{Approach Ball}. If the latter ends successfully, the fallback node returns \textit{Success}. If the condition checking returns \textit{Success}, the fallback node returns success, and the tick is then propagated to the \textit{Grasp Ball} action node. If the latter returns \textit{Success}, the sequence node in the subtree returns \textit{Success} as well, and the tick is finally propagated to the \textit{Place Ball} action node. Note that, should any of the nodes associated with the sequence tasks return \textit{Running}, the root sequence node returns \textit{Running} as well. For this reason, the tree is ticked repeatedly to allow its complete execution.

\subsection{State Machines}
The main concept of the state machine model is to describe a complex system's behavior through states and events. 
In robotics, state machines have become a common choice for modeling task-level control and reactive behaviors, as they offer a way to specify how a robot should respond to internal or external events~\cite{bohren2010smach}.
Given the typical complexity of robotic systems, with numerous states and events, an \sm  must be structured in a modular and hierarchical way to avoid unstructured or chaotic models~\cite{harel1987statecharts,rafconBrunnerSBD16}.

Although many semantics are available for state machines, the most commonly referenced is the one defined in the UML standard~\cite{omg_uml_2017}. Therefore, we present the core concepts of state machines as outlined in the standard.

\begin{figure}[!t]
    \centering
    \includegraphics[width=0.45\textwidth]{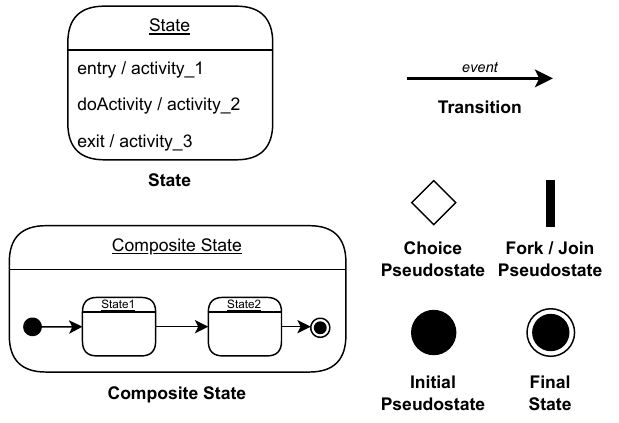}
    \caption{Core \sm\ elements.}
    \label{fig:sm_elements}
\end{figure}

A \textit{State} is a situation in the \sm\ where a specific constraint is maintained. While in this state, activities linked to its status can be carried out. Specifically, a state may include an \textit{entry} behavior executed upon entering the state and an \textit{exit} behavior executed upon leaving the state. Additionally, it can include a \textit{doActivity} behavior that begins after the completion of the entry behavior. 
If a state contains substates, it is referred to as a \textit{Composite State}, allowing the definition of a hierarchical structure among states. States are linked by \textit{Transitions}, labeled with events that trigger the transition between states.
Moreover, the standard adopts pseudostates to abstract different types of elements that define the transition flow. Among them, a \textit{Choice} represents a conditional decision, where the behavior is constrained by evaluating the transition guards associated with the pseudostate. Differently, a \textit{Fork} and \textit{Join} pseudostates serve to split or join multiple transitions.
Finally, the activation and completion of a behavior is regulated by an \textit{Initial} pseudostate representing the starting point and by a \textit{Final} state representing the ending of the behavior.
The visual representation of the core elements is depicted in Fig.~\ref{fig:sm_elements}.

\begin{figure}[!t]
    \centering
    \includegraphics[width=0.95\linewidth]{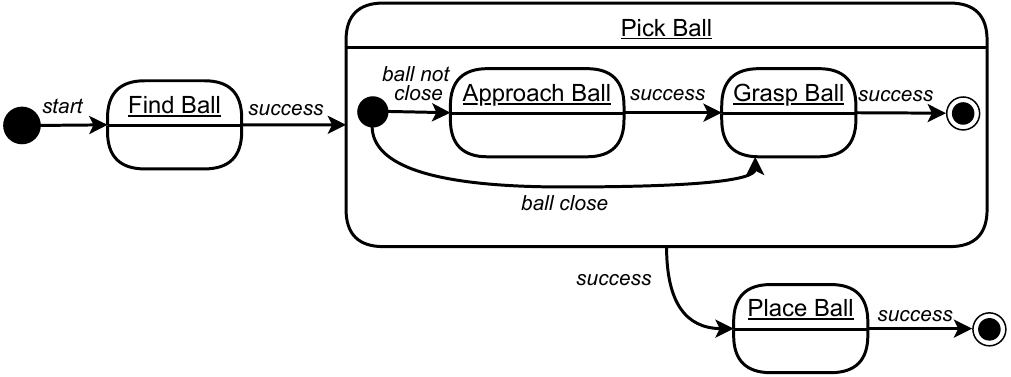}
    \caption{Example of mission expressed as a \sm.}
    \label{fig:sm_example}
\end{figure}

The execution is \textit{event-driven}, the \sm\ is traversed from the initial to the final state based on the triggered transition.
For instance, in the mission example shown in Figure~\ref{fig:sm_example}, the state machine begins by triggering the \textit{Find Ball} state. Upon successful completion of this state, the \textit{Pick Ball} composite state is activated. Within this composite state, the \textit{Approach Ball} state is executed only if the ball is not already near the robot; otherwise, the \textit{Grasp Ball} state is executed directly. If the composite state completes successfully, the \textit{Place Ball} state is activated. Once this state achieves a successful outcome, the final state is reached.

\subsection{Hierarchical Task Networks}

Hierarchical Task Networks (\htns{}) are an automated planning formalism in which high-level tasks are decomposed into progressively simpler subtasks until executable primitive actions are reached~\cite{erol1994htn}. An HTN planning problem starts with an initial task network consisting of tasks and constraints. Primitive tasks correspond to actions that can be directly executed, while non-primitive (compound) tasks must be refined using methods, predefined decompositions into subtasks that preserve ordering and constraint relationships. Planning proceeds by repeatedly selecting non-primitive tasks and replacing them with their corresponding task networks from applicable methods until only primitive tasks remain; a successful plan is then a fully expanded, executable sequence of primitive actions that satisfies all task constraints. 

\htns{} formalize \emph{planning problems} for robot missions by representing a mission as a \emph{task network} that must be refined into an executable course of action. An HTN model distinguishes \emph{compound} (non-primitive) tasks from \emph{primitive} tasks (actions). Compound tasks are refined via \emph{methods}, i.e., domain-defined decomposition rules that replace a task with a (partially ordered) network of subtasks, optionally subject to ordering and state constraints. Planning proceeds by repeatedly selecting a compound task and applying an applicable method until a \emph{primitive task network} is obtained; a plan is valid if the resulting primitive actions can be linearized to satisfy all ordering/causal constraints and are executable in the world state \cite{erol1994htn}. In robotics, HTNs are typically used as a \emph{deliberative} mechanism: they encode domain knowledge to generate structured mission plans before execution, and they support controlled refinement and repair during execution when conditions change. This role aligns with distributed deliberative architectures for multi-robot missions, where an offline-computed hierarchical plan is executed by local supervisors that manage their allocated plan parts and perform hierarchical repair to handle failures while reducing communication demands \cite{lesire2016distributed}.

There is no universally accepted graphical standard for \htns{}. In this work, we adopt a graphical notation consistent with prior robotics literature, where tasks are represented as nodes and hierarchical decomposition is shown by connecting compound tasks to their subtasks through method links, often annotated with ordering constraints or guard conditions~\cite{lesire2016distributed,rodrigues2022architecture,filippone2024handling}. This notation distinguishes between compound tasks (which decompose into subtasks) and primitive tasks (which correspond to executable steps), supporting intuitive interpretation of task hierarchies and dependencies. Notably, in~\cite{lesire2016distributed,rodrigues2022architecture,filippone2024handling} compound tasks are referenced to as \textit{abstract} tasks, while primitive tasks are referenced to as \textit{elementary} or \textit{concrete}. In this paper, we use terms \textit{compound} and \textit{primitive}, consistently with HTN literature~\cite{ghallab2016automated,georgievski2014overview,holler2020hddl}.

\begin{figure}[!t]
    \centering
    \includegraphics[width=\linewidth]{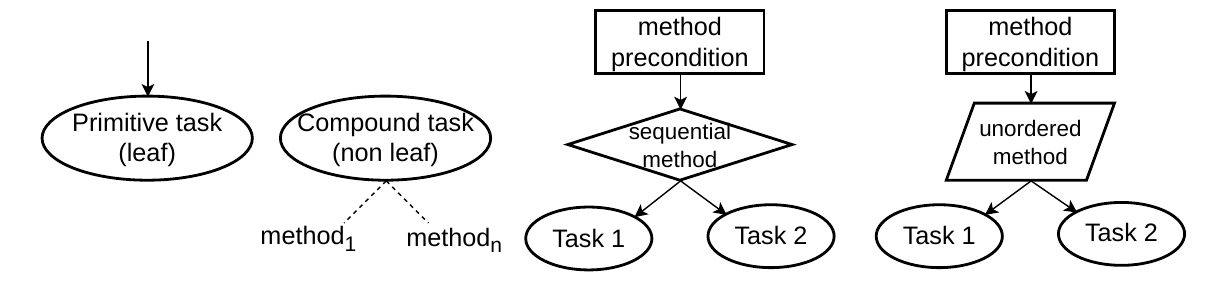}
    \caption{Core \htn\ elements.}
    \label{fig:htn_elements}
\end{figure}

Fig.~\ref{fig:htn_elements} reports the graphical representation of \htn elements we use in this work. Tasks are represented through ellipses. Primitive tasks are the leaves in the tree, while compound tasks are the internal nodes. Methods are depicted according to their execution order: diamonds represent sequential ordering, while parallelograms represent unordered relations of their subtasks. Sequential ordering prescribes tasks that have to be executed in sequence (in the graphical representation, from the left-most to the right-most task). Unordered methods can be performed in parallel (depending on the execution platform, e.g., if assigned to different robots) or in any sequential order. Method can be optionally guarded by preconditions, reported into a rectangular box. They are required to select the proper method to realize the compound task decomposition according to the current conditions.

\begin{figure}[!t]
    \centering
    \includegraphics[width=0.84\linewidth]{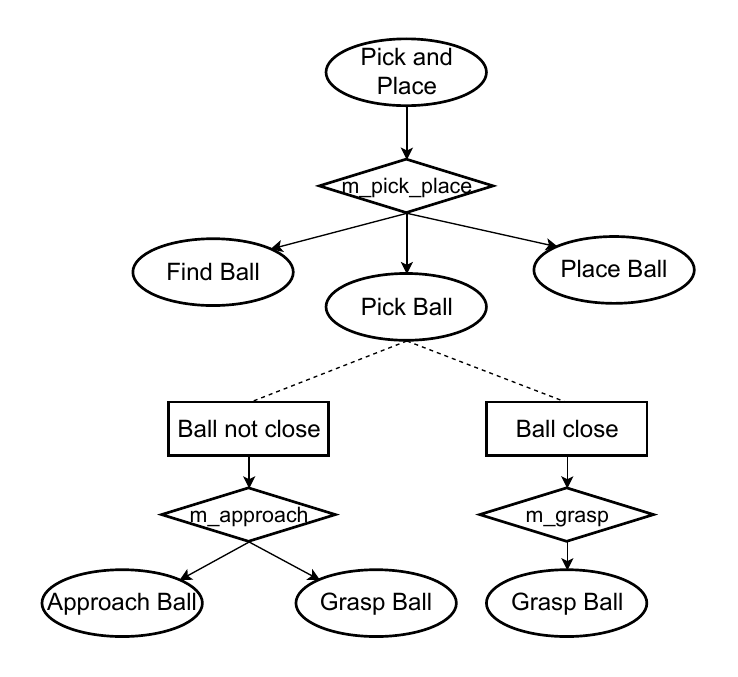}
    \caption{Example of mission expressed as a \htn.}
    \label{fig:htn_example}
\end{figure}

Referring to the example Pick and Place mission, as shown in Fig.~\ref{fig:htn_example}, it is represented in \htn as a compound task that is decomposed through a method (\textit{m\_pick\_place}) into subtasks such as Find Ball, Pick Ball, and Place Ball. Subtasks may themselves be compound and further refined using alternative methods, as shown for Pick Ball, which branches depending on contextual conditions (e.g., whether the ball is close). These conditions guide the selection of appropriate methods (e.g., \textit{m\_approach} or \textit{m\_grasp}), ultimately yielding primitive tasks such as Approach Ball and Grasp Ball that can be directly executed by the robot. This hierarchical decomposition explicitly captures task structure, decision points, and execution dependencies, making HTNs well-suited for modeling complex, goal-directed robotic missions.

\subsection{Business Process Model and Notation}

Business Process Management (BPM) is a discipline widely adopted by organizations to ensure consistent outcomes and identify improvement opportunities~\cite{dumas2018fundamentals}. In particular, BPM manages the chain of events, activities, and decisions connected to an organization. These chains are represented as business process models, and BPM includes concepts, methods, and techniques to support their design, enactment, monitoring, and analysis~\cite{Weske19}.
Over the last years, with the widespread of autonomous and interconnected devices, novel solutions have been focused on applying BPM techniques to specify and drive also robotic missions~\cite{de2017mission,FaMe}.

Business process models are mostly expressed via the \bpmn\ standard~\cite{omg_business_2011}, which is adopted by industry and academia to enable clear, versatile representations for business users and technical developers.
To foster the usage and interchangeability of \bpmn\ between different tools, the models can be shared in a standard manner. Indeed, the standard defines a unique XML-based notation in which a business process is described in a tree-structured way, bringing all the information required for reproducing the elements composing the diagram. Indeed, each \bpmn\ element can be mapped to an XML fragment containing semantic and visual information.
\bpmn\ allows the design of different kinds of diagrams: process, collaboration, choreography, and conversation diagrams. Specifically, \textit{collaboration diagrams} can be used to depict processes in a distributed system. Within these diagrams, various \bpmn\ elements are utilized to model the intended behavior of the referenced system.
Notably, the \bpmn standard defines more than 200 distinct elements~\cite{CompagnucciCFR24}, providing a highly expressive and structured notation for designing collaborative processes.
In the following, we describe the core concepts of \bpmn\ elements. Given the richness of the notation, we provide an abstract overview of these elements rather than an exhaustive description.

\begin{figure}[t]
    \centering
    \includegraphics[width=0.35\textwidth]{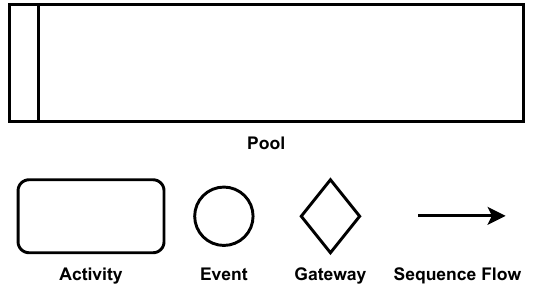}
    \caption{Core \bpmn\ elements.}
    \label{fig:bpmn_elements}
\end{figure}

\textit{Pools} are used to represent participants or organizations involved in the collaboration and include details on internal process specifications and related elements. \textit{Activities} are used to represent a specific work to be performed within a process. \textit{Events} are used to represent something that can happen. \textit{Gateways} are used to manage the flow of a process. Notably, activities, events, and gateways can be marked in different ways to indicate the corresponding execution behavior (e.g., a cross symbol in a gateway marks an exclusive choice). Finally, \textit{Sequence Flows} are used to specify the internal flow of the process, thus the execution order of elements in the same pool.
The visual representation of the core elements is shown in Fig.~\ref{fig:bpmn_elements}.

\begin{figure}[b]
    \centering
    \includegraphics[width=0.48\textwidth]{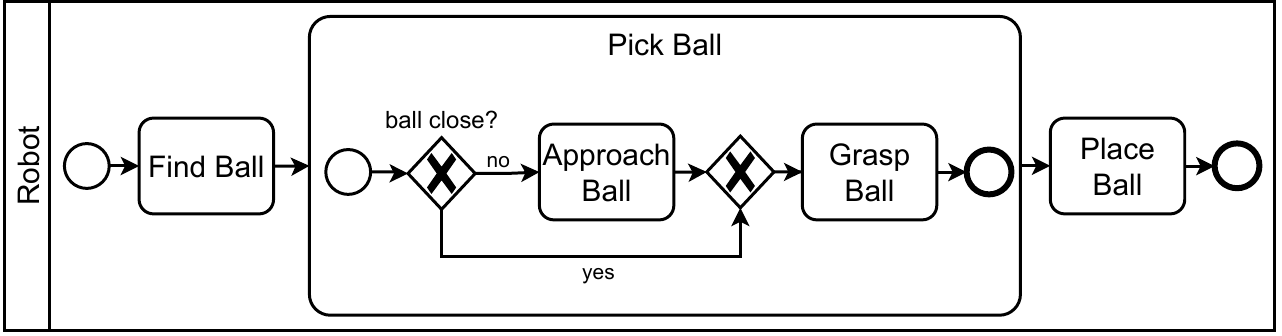}
    \caption{Example of mission expressed in \bpmn.}
    \label{fig:bpmn_example}
\end{figure}

The execution semantics of \bpmn\ is \textit{token-based}~\cite[Sec.~7.1.1]{omg_business_2011}. A token traverses, from a start event, the sequence edges of the process and passes through its elements, enabling their execution, and finally, an end event consumes it when it terminates. Process elements acquire one or more tokens from incoming sequence flows for execution. Once finished, they may produce one or more tokens on outgoing sequence flows, depending on their behavior. 
Considering the example in Fig.~\ref{fig:bpmn_example}, the diagram contains one pool, named \textit{Robot}. The execution starts with one token in the \textit{start event} which traverses sequentially the process model. This activates the first activity (i.e., \textit{Find Ball}) followed by the \textit{Pick Ball} subprocess. The subprocess executes the \textit{Approach Ball} task only if the \textit{ball close} condition is evaluated as false, after that it executes the \textit{Grasp Ball} task. The execution continues by firing the \textit{Place Ball} task and completes when the token reaches the \textit{end event}.

Following the execution semantics, \bpmn\ process models can be directly executed by \bpmn\ engines. These engines can consume and execute processes provided in the correct format. Notably, standardization of the format and semantics by \bpmn\ ensures that the execution behavior remains consistent across different engines~\cite{geigerHLCVW15}.

\section{Research Method and Analysis Corpus}\label{sec:methodology}

This section describes the research method adopted in our study and the analysis corpus used to compare the considered formalisms. Specifically, we outline the sources and materials that informed our analysis, including (i) primary documentation and representative applications of the formalisms in robotics, (ii) a set of robotic scenarios spanning multiple domains, and (iii) a collection of publicly available tools that support, to varying extents, the four formalisms. Together, these elements provide the empirical basis for a systematic and practice-oriented comparison.

\begin{figure}[!t]
    \includegraphics[width=\linewidth]{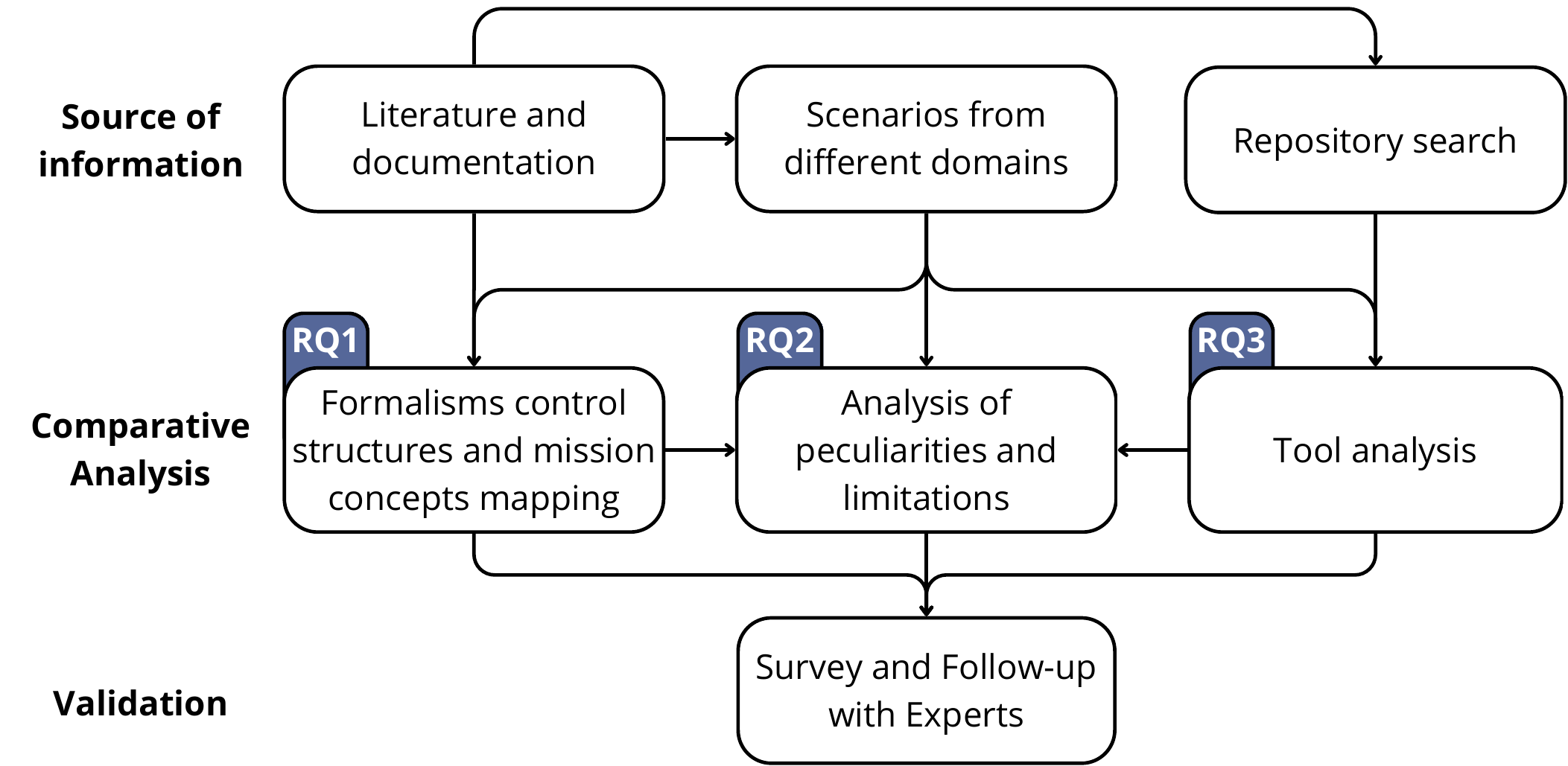}
    \caption{Research method.}
    \label{fig:methodology}
\end{figure}

\Cref{fig:methodology} summarizes our research method for addressing the three RQs, organized into three phases: data collection, comparative analysis, and validation of the results.

\subsection{Sources of Information}\label{sub:methodology-sources}

Our study started by gathering background knowledge on the four considered formalisms.

To this purpose, we collected sources through a targeted literature search on Scopus\footnote{\url{https://www.scopus.com}}, focusing on each of the four formalisms.
A search string was composed for each formalism, using keyword combinations including the formalism name and common variants (e.g., ``hierarchical task network'', ``HTN'', etc.) along with robotics- and mission-related terms.
Following this strategy, we obtained four search strings following this pattern: \textit{``($<$formalism$>$) AND (robot OR robotic) AND (mission OR mission specification OR mission execution)''}, where \textit{$<$formalism$>$} was substituted with the string composed by the name of the formalism and its variants.
The obtained results were filtered according to the following inclusion criteria: (i) the paper focuses on the use of the formalism for robotic mission specification or execution; and (ii) it presents, applies, or discusses the formalism in a robotic context. As exclusion criteria, we discarded works in which the formalism was used for purposes different from mission modeling/execution (e.g., modeling physical space or state machines in control-theoretic contexts).
After the paper filtering, we applied snowballing to get additional potential sources.
The resulting set of sources consisted of papers specifically discussing the properties and applications of \bts~\cite{colledanchise2018behavior,colledanchise2016advantages,colledanchise2021on,ghzouli2020behavior,iovino2022survey,GUGLIERMO2024104714} and \sms~\cite{thorsten2023behavior} in robotics, comparison between \bts and \sms~\cite{dragule2025effects,IovinoFFCSS25}, applications of \htn in robotics~\cite{lesire2016distributed,filippone2024handling,gil2023mission,rodrigues2022architecture}, and applications of \bpmn in robotics~\cite{FaMe,whitaker2024mission,de2017mission,de2020event}.
Moreover, we looked for sources that document the formalism, regardless of the application domain, including scientific papers~\cite{erol1994htn}, informal documentation, and standard definitions~\cite{omg_business_2011,omg_uml_2017}.

The obtained sources were further exploited to identify (i) a set of robotic scenarios spanning multiple domains (e.g., logistics, healthcare, households, and agriculture), and (ii) a set of publicly available tools that support, to different extents, the four formalisms.

Concerning the identification of the scenarios, besides the aforementioned sources, we also considered the RoboMAX exemplars collection~\cite{robomax}.
The scenarios were selected by applying the following inclusion criteria: (i) the missions in the scenarios should involve multi-purpose robot capabilities rather than fixed, single-purpose behaviors, and (ii) missions take place in dynamic environments, thereby excluding, for example, single-purpose industrial robots. As exclusion criteria, we did not include simple missions concerning fixed sequences of tasks (e.g., pick and place). 
In total, we identified 11 scenarios, representing robot behavior at different abstraction levels over different domains.
In particular, we considered the \textit{Pick Ball}~\cite[Example 2.1]{colledanchise2018behavior} and \textit{Humanoid Robot}~\cite[Figure 2.4]{colledanchise2018behavior} scenarios, the \textit{Vital Signs Monitoring}, \textit{Keeping Clean}, \textit{Food Logistics}, \textit{Lab Samples Logistics}, \textit{Welcome People to Hospital}, and \textit{Deliver Goods} exemplars from RoboMAX~\cite{robomax}, the \textit{SUAVE} use case from~\cite{suave}, the \textit{Smart Agriculture} use case from~\cite{FaMe}, and the \textit{Warehouse Automation} scenario from~\cite{ReyCCMC19}.
The detailed description of each scenario, as well as the missions modeled using the formalisms, is available in the dedicated section of the replication package~\cite{replicationpackage}.

Finally, the selection of the tools supporting the formalisms was performed by leveraging the ones mentioned, analyzed, or used in the selected information sources.
Additionally, we scoured GitHub repositories and reviewed tools mentioned in the literature, grouping them by purpose. To this aim, we searched for different strings containing the formalism name (both in full and in acronym) and ``robot'' or ``robotics'' (e.g., ``bt robot'', ``bt robotics'', ``behavior tree robot'', etc.). 
As inclusion criteria, we considered (i) tools designed specifically for robotic missions, and (ii) general-purpose tools for the considered formalisms that can be adapted to robotics. As exclusion criteria, we considered (i)~lack of publicly available documentation, (ii)~educational or prototype implementations used as toy examples, (iii)~not maintained tools, i.e., last commit older than three years\footnote{The search was done in July 2025.}.

\subsection{Comparative Analysis}\label{sub:methodology-analysis}
Starting from the literature and the documentation of the formalisms, we analyzed the formalisms based on how they support the modeling of (i) the control structures for describing the flow of actions to be performed in the robotic mission, and (ii) the main concepts related to robotic missions, identified from the literature and within the scenarios.
The result of the analysis conducted for \textit{RQ1} consists of a mapping of the base elements offered by the formalisms to the aforementioned control structures and robotic mission concepts. Results are presented in \Cref{sec:concepts}.

By leveraging the results of the analysis performed for \textit{RQ1} and the existing literature, we analyze the peculiarities and limitations of formalisms in modeling robotic missions with respect to mission-level concerns (\textit{RQ2}).
These concerns were derived from the identified scenarios and capture recurring modeling needs emerging from the missions to be specified, such as the need for specifying reactive behavior, decision making, interactions with humans or other robots, waiting, and time-dependent behavior.
To identify them, we inspected the 11 scenarios and recorded the recurring modeling needs that affected mission specification. A concern was selected if it respected the following criteria: (i) it appears as a mission-level requirement, and (ii) it affects the mission specification rather than the low-level implementation of actions. The resulting set was then consolidated during the modeling and review process described below.
Then, we modeled the mission of each of the 11 identified scenarios using the four formalisms, employing existing straightforward tools when available, and evaluated the resulting models.
Each scenario was modeled by keeping the same abstraction level as in the scenario description. We considered as base actions (i.e., skills) the ones that are reported in the scenario description. We leveraged the control structures and concept modeling identified in the scope of \textit{RQ1} and the modeling tools that were previously selected to stress them in the modeling of complex behaviors.
The modeled scenarios allowed us to analyze the formalisms' expressiveness, by scoring their suitability in expressing the identified mission concerns.

The analysis process was realized according to the following methodology:
\begin{enumerate}
    \item two co-authors modeled independently and separately different scenarios using the different formalisms;
    \item the models obtained by one of the two co-authors were reviewed by the other to check for model correctness, and vice versa;
    \item a third co-author reviewed the models and facilitated the discussion for the identification of the model characteristics.
\end{enumerate}
Steps (1) and (2) were essential, as, similarly to software programming, there is no single, uniquely correct way to model a scenario. Multiple valid representations may exist, and this process allowed us to cross-check the soundness of the models while mitigating individual modeling biases.
The output of the analysis conducted for \textit{RQ2} allowed highlighting the strengths and weaknesses of each of the formalisms in modeling different aspects of robotic missions. The results of the analysis are reported in \Cref{sec:expressivity}.

Finally, we analyzed the tools associated with the formalisms by considering their scope and their usability in robotic missions, focusing on those that are actively maintained. We also examined the baseline tools that support each formalism and have served as the foundation for the development of current ROS-compatible packages.
The tool analysis allowed us to (i) support the results conducted for \textit{RQ2}, in particular concerning the analysis of the strengths and weaknesses of the formalisms, as some of the tools provide implementation-level solutions for expressing robotic mission concerns, and (ii) draw an overview of the major currently available tools supporting the formalisms within the robotic domain.

\subsection{Validation}\label{sub:methodology-validation}

To validate the findings derived from our analysis, we conducted questionnaire surveys with domain experts. For each research question, the experts evaluated our results in terms of \textit{completeness}, \textit{correctness}, and \textit{alignment} with established formalisms and best practices for their use in robotics.

We followed the \textit{questionnaire surveys empirical standard} and its essential attributes~\cite{acmsigsoft}. Specifically, this standard prescribes the systematic collection of data from a defined sample of participants through a structured set of questions, typically managed via computerized forms.
Participants were selected through \textit{purposive sampling}~\cite{etikan2016comparison}, focusing on authors of the scientific works that we used as references for this study, as they have direct expertise in the corresponding formalisms and their application to robotic systems. Each expert was contacted via email and received timely reminders to encourage participation.

The questionnaire was custom-designed to facilitate targeted and reliable evaluation. To reduce respondent burden and ensure relevance, each participant was asked to declare their expertise on each of the formalisms by rating the expertise using a 5-point self-assessment scale (1 = not familiar, 2 = heard of it, 3 = some experience, 4 = used many times, 5 = expert user). Participants were asked to reply only to the questions related to the formalism(s) in which they had acknowledged expertise (i.e., expertise higher than or equal to 3).
The survey primarily consisted of \textit{close-ended Likert-type questions} (1 = strongly disagree to 5 = strongly agree), assessing agreement with statements about the completeness, correctness, and alignment with our findings.
To strengthen results interpretability, every closed-ended question was complemented with an open text field, mandatory for responses rated less than or equal to 3 (neutral or lower), requiring participants to justify their assessment. This design choice ensured that lower evaluations were always supported with qualitative explanations.

Responses were collected in a structured spreadsheet for a posteriori analysis.
Following the analysis of the questionnaire responses and an internal discussion among the authors, we complemented the survey with a round of in-depth follow-up interactions.
Participants to the follow-up round were selected based on: (i) their declared level of expertise, ensuring at least one self-reported 5/5 expert for each formalism; (ii) their expertise span multiple formalisms; (iii) the presence of particularly critical or insightful questionnaire responses; (iv) have background knowledge of the formalism, also beyond robotic applications; and (v) have acknowledged the willingness to be contacted for follow-up questions.

Questionnaire data collection was carried out over a period of four weeks in January 2026, resulting in a total of 29 complete responses out of 83 invitations, corresponding to a response rate of 34.94\%. Table~\ref{tab:interviews} overviews the questionnaire participants profiles.
The respondent group comprised experts across different profiles and career stages, including 4 PhD students, 6 postdoctoral researchers, 3 researchers, 11 professors, and 5 roboticists and industry people.
Regarding expertise on the formalisms, participants self-reported experience as follows: 27 experts in SMs, 22 in BTs, 14 in BPMN, and 10 in HTN, with an average experience of $\sim$8 years in robotic software engineering.

Follow-up interactions were carried out over two weeks in February 2026. Seven participants were invited for follow-up discussions, of whom four confirmed their availability.
Specifically, we collected detailed feedback from four participants, whose profiles are highlighted in Table~\ref{tab:interviews}.
Feedback was collected both synchronously and asynchronously, according to participant availability. Three participants (\textit{[par:5]}, \textit{[par:10]}, and \textit{[par:17]}) were interviewed in live sessions, while one (\textit{[par:20]}) provided written responses to a set of follow-up questions.
The purpose of this extended feedback collection was to deepen the discussion of the comparative analysis, focusing on critical points and clarification requests raised in the questionnaire, as well as on additional issues that emerged from participants' comments and required further rationale or refinement.
Therefore, the results reported in \Cref{sec:concepts}, \Cref{sec:expressivity}, and \Cref{sec:tools} reflect the refinements made to the findings after analyzing the feedback collected through the expert questionnaire and follow-up interactions. The corresponding validation subsections explain how participants' feedback informed revisions, extensions, and clarifications of the comparative analysis.

\begin{table}[t]
    \centering
    \caption{Questionnaire participants overview (grey-highlighted rows indicate interviewed participants).}
    \resizebox{\linewidth}{!}{
    \begin{tabular}{llp{1cm} cccc}
        \toprule
         \multirow{2}{*}{\textbf{ID}} & \multirow{2}{*}{\textbf{Profile}} & \multirow{2}{*}{\textbf{Years}} & \multicolumn{4}{c}{\textbf{Declared expertise}} \\
         \cmidrule{4-7}
         &&\textbf{in role}& \textbf{BT} & \textbf{SM} & \textbf{HTN} & \textbf{BPMN} \\
         \midrule
         par:1 & Postdoc & 1 & 2/5 & 4/5 & 1/5 & 5/5 \\
         par:2 & Postdoc & 5 & 5/5 & 5/5 & 3/5 & 3/5 \\
         par:3 & Industry & 4 & 4/5 & 3/5 & 5/5 & 3/5 \\
         par:4 & PhD student & 5 & 5/5 & 5/5 & 5/5 & 3/5 \\
         \rowcolor{mygray} par:5 & Roboticist & 7 & 5/5 & 4/5 & 3/5 & 2/5 \\
         par:6 & Professor & 13 & 4/5 & 4/5 & 2/5 & 1/5 \\
         par:7 & Professor & 18 & 3/5 & 4/5 & 5/5 & 2/5 \\
         par:8 & Postdoc & 7 & 3/5 & 3/5 & 3/5 & 5/5 \\
         par:9 & Professor & 20 & 4/5 & 4/5 & 2/5 & 1/5 \\
         \rowcolor{mygray} par:10 & Professor & 3 & 3/5 & 5/5 & 1/5 & 4/5 \\
         par:11 & Professor & 10 & 2/5 & 3/5 & 5/5 & 2/5 \\
         par:12 & Postdoc & 8 & 5/5 & 4/5 & 2/5 & 1/5 \\
         par:13 & Professor & 5 & 3/5 & 5/5 & 4/5 & 5/5 \\
         par:14 & PhD student & 6 & 5/5 & 3/5 & 2/5 & 2/5 \\
         par:15 & Researcher & 7 & 4/5 & 5/5 & 1/5 & 1/5 \\
         par:16 & Researcher & 3 & 3/5 & 3/5 & 1/5 & 5/5 \\
         \rowcolor{mygray} par:17 & Professor & 3 & 2/5 & 2/5 & 1/5 & 5/5 \\
         par:18 & Professor & 26 & 3/5 & 5/5 & 1/5 & 1/5 \\
         par:19 & Postdoc & 5 & 2/5 & 3/5 & 2/5 & 2/5 \\
         \rowcolor{mygray} par:20 & Researcher & 3 & 2/5 & 5/5 & 2/5 & 2/5 \\
         par:21 & Postdoc & 3 & 3/5 & 4/5 & 4/5 & 5/5 \\
         par:22 & Professor & 4 & 1/5 & 2/5 & 1/5 & 5/5 \\
         par:23 & PhD student & 4 & 5/5 & 3/5 & 1/5 & 2/5 \\
         par:24 & Professor & 21 & 5/5 & 5/5 & 2/5 & 2/5 \\
         par:25 & Industry & 15 & 3/5 & 5/5 & 2/5 & 1/5 \\
         par:26 & Professor & 10 & 3/5 & 5/5 & 2/5 & 3/5 \\
         par:27 & PhD student & 6 & 3/5 & 4/5 & 1/5 & 3/5 \\
         par:28 & Roboticist & 5 & 4/5 & 4/5 & 3/5 & 1/5 \\
         par:29 & Roboticist & 11 & 2/5 & 4/5 & 2/5 & 5/5 \\
        \bottomrule
    \end{tabular}
    }
    \label{tab:interviews}
\end{table}

The questionnaire is available online in the replication package~\cite{replicationpackage}. Supplementary material also includes the anonymized answers and the interview summary.

\section{Control Structures and Mission Concepts (RQ1)}\label{sec:concepts}

This section addresses \textit{RQ1} by analyzing how the considered formalisms model control structures and mission concepts. Following the first phase of our research method (\Cref{sec:methodology}), we draw on (i) insights from primary documentation and representative robotics applications, and (ii) a set of robotic scenarios spanning multiple domains. Grounding the analysis in both language constructs and concrete mission scenarios enables a systematic assessment of how each formalism represents mission control flow and core mission abstractions.

\begin{table*}[!htb]
\caption{Control structures.}
\begin{tabular}{l| p{0.21\textwidth} p{0.21\textwidth} p{0.21\textwidth} p{0.21\textwidth}}
\toprule
\textbf{} & \textbf{Sequential} & \textbf{Parallel} & \textbf{Conditional} & \textbf{Loop} \\ \midrule
\multirow{3}{*}{\textbf{BT}} & 
From left to right from the Sequence or Fallback node & Parallel Node & Combination of Fallback, Sequence, and Condition Nodes & Decorator (possible implementation)  \\ 
& \includegraphics[width=0.67\linewidth]{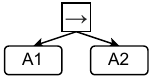} &
\includegraphics[width=0.67\linewidth]{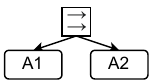} &
\includegraphics[width=0.9\linewidth]{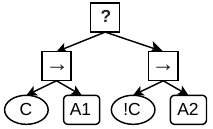} &
\includegraphics[width=0.57\linewidth]{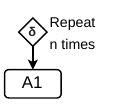}
\\
& $A1$ is executed before $A2$
& $A1$ and $A2$ are executed in parallel 
& If $C$ is true, then $A1$ is performed; if $C$ is false, then $A2$ is performed
& $A1$ is executed $n$-times 
\\
\midrule
\multirow{3}{*}{\textbf{SM}} & 
From initial to final state following triggered events & Fork pseudostate & Choice pseudostate & Transition cycles \\
& \includegraphics[width=0.75\linewidth]{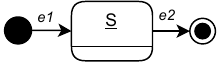}  
& \includegraphics[width=0.9\linewidth]{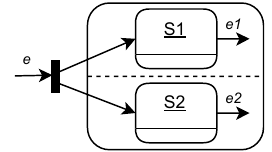} 
& \includegraphics[width=.95\linewidth]{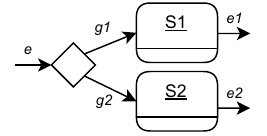}
& \includegraphics[width=0.7\linewidth]{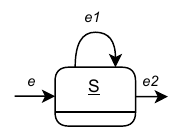} \\
& When $e1$ is triggered, state $S$ remains active until $e2$ is triggered 
& The fork pseudostate splits the incoming transition into two transitions, activating $S1$ and $S2$ 
& If $g1$ is true, $S1$ is executed; otherwise if $g2$ is true, $S2$ is executed & $e1$ reactivates $S$'s action, while $e2$ terminates it
\\
\midrule
\multirow{3}{*}{\textbf{HTN}} 
& Method with sequential relationship
& Method with unordered relation 
& Methods combination with different preconditions
& Achievable using recursive methods and conditions \\
& \includegraphics[width=0.65\linewidth]{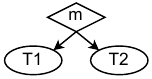} 
& \includegraphics[width=0.65\linewidth]{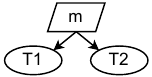} 
& \includegraphics[width=0.57\linewidth]{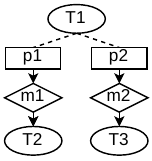} 
& \includegraphics[width=0.8\linewidth]{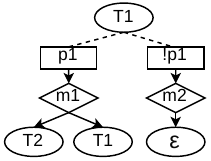} \\
& $T1$ is executed before $T2$
& $T1$ and $T2$ are executed in any order, in parallel if possible
& If $m1$'s preconditions hold ($p1$), $T2$ is executed; if $m2$'s preconditions hold ($p2$), $T3$ is executed
& As long as loop preconditions ($p1$) hold, $T2$ is executed, then $T1$ recursively runs the loop; when $p1$ does not hold anymore, $m2$ realizes the loop exit
\\
\midrule
\multirow{3}{*}{\textbf{BPMN}} 
& From the start to the end event following the sequence flow 
& AND Gateway 
& XOR Gateway  
& $(a)$ Combination of XOR gateways; $(b)$ Loop Activity; $(c)$ Multi-instance Activity \\
& \includegraphics[width=\linewidth]{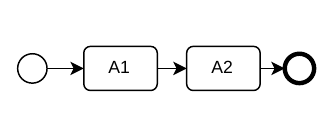} 
& \includegraphics[width=0.8\linewidth]{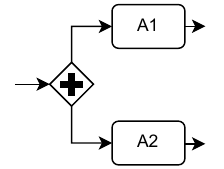} 
& \includegraphics[width=0.9\linewidth]{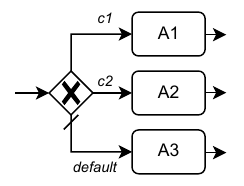} 
& \includegraphics[width=\linewidth]{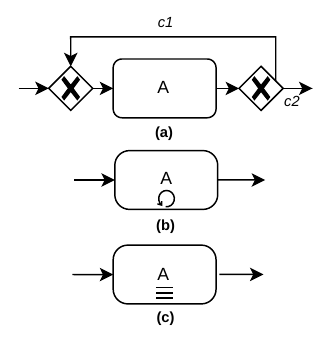} \\
& The flow transitions to activity $A1$, then to $A2$, after which the process terminates.
& $A1$ and $A2$ are executed in parallel
& If $c1$ is true, $A1$ is executed; if $c2$ is true, $A2$ is executed; if neither $c1$ nor $c2$ is true, the default flow is taken and $A3$ is executed
& $(a)$ 
$A$ is executed repeatedly until condition $c1$ is no longer true 
$(b)$ $A$ 
is executed until a specified condition is met 
$(c)$ $A$ is sequentially executed for a given number of times
\\
\bottomrule
\end{tabular}
\label{tab:control-flow-op}
\end{table*}

\subsection{Control Structures}

Our analysis of control flow considers \textit{sequential}, \textit{conditional}, and \textit{loop} constructs, which are not specific to robotics and originate from structured programming and flowchart principles~\cite{nassi1973flowchart}.
They build the fundamental blocks for control flow both in most general-purpose programming languages and many different modeling frameworks, e.g., UML activity diagrams, business process models, process algebra, including the ones considered in this paper. 
Additionally, we consider the \textit{parallel} control structure, since both single multi-purpose robots and multi-robot missions commonly require actions to be performed concurrently, and parallelism is a widely adopted construct across the aforementioned modeling frameworks.
Table~\ref{tab:control-flow-op} presents how the formalisms support control structures that drive the flow of a mission. Additionally, we provide a graphical representation of these control structures to exemplify their functionality.

{\bf Sequential}: To realize a sequential task execution, \bt{}s employ the \textit{sequence} and \textit{fallback} control flow nodes. These nodes demand the execution (i.e., \textit{ticking}) of their children from left to right interrupting the sequence when a child returns \textit{Failure} (sequence nodes) or \textit{Success} (fallback nodes). 
In contrast, \sms{} do not have an explicit control structure to model a sequence of actions; rather, an \sm\ transitions from the initial to the final state, reacting dynamically to triggered events as the system evolves, i.e., the sequence is driven by the events that trigger state changes. In \htn\, task sequences are realized through sequential methods, where all the method's children are executed from left to right. Finally, a \bpmn\ is traversed from the start event to the event node, based on the sequence flow.

{\bf Parallel}: Concurrent execution flow is necessary to model parallel behaviors. \bt\ offers the parallel node to compose child nodes that must be executed concurrently. In a \sm, a fork pseudostate can be used to split the incoming transition into multiple transitions, without guards, activating the corresponding states. \htn\ offers \textit{unordered methods}, where all the children can be executed in any order, even in parallel if possible. It is worth noting that this type of method does not explicitly demand or constrain parallel execution of tasks; rather, the control architecture that executes the task is responsible for managing their parallel execution. In \bpmn, the AND gateway receives an incoming token and splits it into multiple tokens for each outgoing flow, thus enabling concurrent flow execution.

\begin{obs}{Execution parallelism}\label{obs:exec-par}
Concurrency cannot always be fully realized in practical implementations, as it depends on the robotic platform and controller implementation. For instance, parallelism across tasks that share the same resources can often be approximated only by interleaving tasks.
\end{obs}

{\bf Conditional}: Modeling a conditional flow is necessary to regulate execution based on specific conditions. In a \bt, this can be achieved by a fallback node with sequence and condition nodes~\cite{colledanchise2018behavior}. Specifically, the condition node's evaluation determines whether to execute the action following the condition node that is evaluated to true.
\begin{obs}{BT nodes \textit{failure} semantics}\label{obs:bt-fail}
A \textit{Failure} state can be either returned by a condition node due to the condition evaluated as \textit{false}, or by an action node due to a failure (e.g., due to errors). To avoid the ``spurious'' execution of actions, \bts require checking both the condition and its negation. This allows \textit{failure} states arising from errors and condition evaluation to be disambiguated.
\end{obs}
An \sm\ utilizes the choice pseudostate to evaluate the guards of outgoing transitions (i.e., $g1$ and $g2$ in Table~\ref{tab:control-flow-op}), determining the subsequent flow of execution. \htn\ does not explicitly model choices and does not have dedicated constructs to evaluate conditions. However, preconditions can be associated with methods that refine compound tasks: different methods can be associated with a compound task, hence using preconditions to specify the conditional behavior to follow. Similar to a \sm, \bpmn\ uses the XOR gateway, which evaluates conditions on sequence flows to determine the direction of execution. Moreover, \bpmn allows the explicit specification of a default flow: if none of the conditions are satisfied, the process follows the default branch; if no default flow is defined, an error is raised.

\begin{obs}{Mutually-exclusive guards}\label{obs:mutually-ex}
Conditional semantics differ across formalisms. \bts resolve simultaneous conditions through implicit prioritization (tick order)~\cite{ghzouli2020behavior}, whereas \sms, and \htns require the modeler to ensure guard mutual exclusivity and exhaustiveness.
In \bpmn, XOR gateways select the first satisfied condition; conditions are evaluated in order, so that the first true condition is considered~\cite[p.435]{omg_business_2011}, while optionally supporting an explicit default flow.
\end{obs}

{\bf Loop}: Finally, iterations allow an execution to be repeated multiple times. In \bt, a \textit{repeat} decorator node can be defined to tick the child node $n$-times or unless the child returns \textit{success}. It is worth remarking that, although being provided by default by the main \bt implementations, such a decorator is not defined within the formalism. However, the behavior of decorators is by definition customizable~\cite{colledanchise2018behavior}, allowing different loop policies to be defined. In an \sm, a transition can loop over a state, keeping it active until the guard in the cycle is triggered. 
In \htn, iterative behavior is not represented through explicit loop constructs but is realized implicitly through recursive task decomposition. In particular, a compound task can be refined by a method whose subtasks include the same compound task, provided that the method's preconditions remain satisfied. The repetition continues as long as these preconditions hold, and terminates when no recursive method is applicable, thereby encoding loop-like behavior through conditional recursion.
In \bpmn, three structures support repetitions. Using XOR gateways, a structured loop repeats the flow inside the gateways as long as the condition remains true. Alternatively, a single activity can be marked as a \textit{loop} and configured to be executed until a given condition is evaluated as true, and can be subject to an optional maximum number of repetitions.
Moreover, BPMN also supports \textit{multi-instance} markers to run a given number of activity instances.
These instances may run sequentially or, when appropriate for the scenario (e.g., dispatching tasks to multiple robots), in parallel.

\subsection{Mission Concepts}

Regarding mission concepts, we adopt the terminology introduced in~\cite{garcia2023software,thorsten2023behavior} and rely on the layered organization proposed in RobMoSys~\cite{robmosys} to structure the representation of robotic capabilities.
\Cref{tab:robmosys} reports such layers, as separate concepts that represent different abstraction levels, each providing a lower-level specification of the concept on top of it.
In line with the RobMoSys abstraction layers, and given our focus on comparing formalisms for high-level mission specification, we do not consider concepts below the \textit{service} layer. These layers address low-level, hardware-dependent execution aspects that are outside the scope of mission-level modeling considered in this study.
Instead, we consider the \textit{skill}, \textit{task} (task plot), and \textit{mission} layers, as follows:

\begin{table}[!b]
\centering
\caption{Abstraction levels in robotic systems (adapted from~\cite{robmosys}).}
\label{tab:robmosys}
\begin{tabular}{lp{5.6cm}}
\toprule
\textbf{Abstraction Level} & \textbf{Example} \\
\midrule
Mission & Serve customers; serve as a butler \\
Task plot & Deliver coffee \\
Skill & Grasp object with constraints \\
Service & Move manipulator \\
Function & Inverse kinematics (IK) solver \\
Execution Control & Activity \\
OS / Middleware & pthread; socket; FIFO scheduler \\
Hardware & Manipulator; laser scanner; CPU architecture; mobile platform \\
\bottomrule
\end{tabular}
\end{table}

\begin{itemize}
    \item A \textit{skill} is a programmed action that represents a basic capability of the robot. Typically, it is implemented by software experts by leveraging the lower-level components, abstracting the implementation details. It provides access to the functionalities realized within the robot's components and makes them accessible to the task level.
    \item A \textit{task} is a symbolic representation of a robotic behavior realized as a combination of skills. Tasks specify {\it what} must be done and only partially {\it how}, abstracting away from the concrete implementations provided by the skills composed to realize them.
    \item A \textit{mission} represents the global high-level objective that the robotic system has to accomplish, defined as a set of coordinated sequences of tasks to be performed that include precedence constraints and that either can be organized in sequences or executed in parallel. Within the scope of this paper, mission constitutes the fundamental element specified using the selected formalisms.
\end{itemize}

Additionally, we consider further concepts that are involved in the mission specification. In particular, we consider the capability of a formalism to express the concepts of \textit{data}, \textit{communication}, \textit{events}, \textit{errors}, and \textit{pre/post-conditions}.
These concepts were derived both from the literature and from the analysis of the scenarios mentioned above, where specific needs naturally emerged.

\begin{itemize}
    \item \textit{Data} specification encompasses the configuration concern~\cite{robmosys} of the system and the management of knowledge propagation throughout the mission. It is required to provide the needed information for skills, tasks, and control structures. Data can be static, provided as an input that configures or parametrizes skills/tasks (e.g., the target location for a navigation task), or dynamic, being produced, managed, and propagated across the skills/tasks performed by robots within a mission (e.g., the status of environmental conditions affecting the mission).
    \item \textit{Communication} in the mission specification is required to address the communication and coordination concerns~\cite{robmosys} of the system, particularly when the mission is defined in a multi-robot context~\cite{FaMe} or when robots must interact with humans or external systems~\cite{ReyCCMC19}. The explicit specification of communication defines how robots share state information (e.g., task execution status) and propagate mission-relevant data (e.g., environmental conditions or context variables).
    \item \textit{Events} enable the definition of how the system responds to internal or external events that arise during mission execution and require explicit management from a mission-level perspective (e.g., executing additional skills/tasks, or reconfiguring them). Handling such events allows for modeling the robot's reactive behavior~\cite{colledanchise2018behavior}.
    \item \textit{Errors} specify the management of a particular class of events that arise from faults, failures, or any unexpected conditions preventing the mission from being executed without proper handling. Explicit error handling allows for the definition of fault-tolerant and resilient behavior~\cite{crestani2015enhancing, colledanchise2018behavior}. 

\begin{table*}[htbp]
\caption{Mission concepts.}
\begin{tabular}{p{0.035\textwidth}p{0.06\textwidth}p{0.07\textwidth}p{0.13\textwidth}p{0.1\textwidth}p{0.13\textwidth}p{0.12\textwidth}p{0.175\textwidth}} 
\toprule
\textbf{} & \textbf{Skill} & \textbf{Task} & \textbf{Data} & \textbf{Communication} & \textbf{Events} & \textbf{Errors} & \textbf{Pre/post-conditions}\\
\midrule
\textbf{\bt} &
Action Node~\cite{colledanchise2018behavior,colledanchise2021on} &
  Sub-tree~\cite{colledanchise2018behavior,colledanchise2021on} &
  Data inputs to action nodes through ports and blackboard storage~\exLink{behaviortree.dev}{https://www.behaviortree.dev/docs/learn-the-basics/xml_format/} \exLink{py-trees.readthedocs.io}{https://py-trees.readthedocs.io/en/devel/blackboards.html} &
  Rely on action implementation~\cite{colledanchise2016advantages} &
  Not explicitly modeled. Achievable thanks to conditions that reactively verify if they are true or not through the reactive nature &
  Not explicitly modeled. Achievable through the reactive nature and \textit{Failure} propagation &
  Not explicitly modeled. Achievable through the Postcondition-Precondition-Action (PPA) pattern~\cite{colledanchise2018behavior} \\
\midrule
\textbf{\sm} &
Simple State~\cite[p.308]{omg_uml_2017} &
  Composite State~\cite[p.308]{omg_uml_2017} &
  Data handled inside states~\cite{omg_uml_2017} and global variables configuration \exLink{flexbe.readthedocs.io}{https://flexbe.readthedocs.io/en/latest/fbetut_1.html} &
  Rely on behavior implementation &
  Each transition is associated with an event. But each transition can only be activated if the related state is active &
  A transition can be related to an error. But each transition can only be activated if the related state is active &
  Transition guards~\cite[p.315]{omg_uml_2017}. A specialization named \textit{Protocol Transition} supports pre- and post-conditions \exLink{uml-diagrams.org}{https://www.uml-diagrams.org/protocol-state-machine-diagrams.html} \\
\midrule
\textbf{\htn} &
Primitive Task &
  Compound Task &
  In task header (e.g., travel(d)) (Or as a parameter in the task definition) &
  Rely on primitive task implementation &
  None &
  None &
  In the task and method definition \\
\midrule
\textbf{\bpmn} &
  Task~\cite[p.154]{omg_business_2011},~\cite{FaMe,schmidbauer2023empirical} &
  Call Activity or Sub-process~\cite[p.430]{omg_business_2011} &
  Data Objects~\cite[p.224]{omg_business_2011} and process variables \exLink{docs.camunda.org}{https://docs.camunda.org/manual/latest/user-guide/process-engine/variables/} &
  Message or Signal Events~\cite[p.269-272]{omg_business_2011},~\cite{FaMe} &
  Multiple event types capable of represent different situations~\cite[p.232]{omg_business_2011},~\cite{de2017mission,FaMe} &
  Error Events~\cite[p.264]{omg_business_2011}~\cite{de2017mission,FaMe} &
  Not natively supported. Intermediate events can be used as a workaround. Some works provide extensions to support these conditions~\cite{IntrigilaPD21,TsaiLW07} \\
  \bottomrule
\end{tabular}
\label{tab:concepts}

\end{table*}
    
    \item \textit{Pre and post-conditions} formalize the states of the system before and after the execution of a task or skill. In particular, \textit{pre-conditions} define the requirements that must hold before executing one or more actions (e.g., the robot has to be in the designed location for picking an object). Instead, \textit{post-conditions} define the expected system or environment state after the successful execution of an action (e.g., the robot holding an object). These explicit definitions allow ensuring the consistency among the sequences of tasks and skills in the mission specification, their dependencies, and enable the support for planning and verification.
\end{itemize}

Table~\ref{tab:concepts} overviews and compares how the formalisms can be used to model the elements a user may need to represent in the robotic mission.
The table also reports the main references for the reported solutions and uses the notation [\textit{$\langle$website$\rangle$}~\faExternalLink{}] to refer to technical or non-academic documentation.
Notably, within the comparison, the \textit{mission} concept is not reported, as for all the formalism we consider the mission as the whole model.

{\bf Skill}: Skills are modeled as atomic elements for all the formalisms we are considering. 
In particular, in \bt{}s, skills can be modeled as leaves in the tree through \textit{action nodes} representing either actuation or sensing operations. In \sms{}, skills can be modeled through \textit{simple states}, which are regulated by the internal state behaviors (i.e., \textit{entry}, \textit{doActivity}, \textit{exit} behaviors). In \htns{}, skills can be modeled through \textit{primitive tasks}. In \bpmn\, they can be modeled through \textit{tasks}, i.e., atomic activities in the notation's standard. Depending on the skill's objective, a task can be classified into different types, such as a service task, which directly calls a robot service~\cite{schmidbauer2023empirical}, or a script task, which embeds robot-specific code within the \bpmn\ task~\cite{FaMe}.

{\bf Task}: 
In \bts{}, tasks are represented by \textit{sub-trees}. In \sms{}, a task can be modeled with \textit{composite states}, which enhance the modularity of the model by nesting simple states enabling task achievement. In \htns{}, they can be represented as \textit{compound tasks}, which are refined into primitive tasks by methods. Finally, in \bpmn\, a task can be modeled through a \textit{call activity} or \textit{sub-process}. The main difference is that a call activity references an external process, while a sub-process is embedded within the original process definition. The primary use case for a call activity is to enable a reusable process definition that can be invoked from multiple other process definitions. For instance, in the example missions in Section~\ref{sec:background}, the task \textit{Pick Ball} is modeled as a subtree in the \bt\ in Figure~\ref{fig:bt_example}, as a composite state in the \sm\ in Figure~\ref{fig:sm_example}, as a compound task in the \htn\ in Figure~\ref{fig:htn_example}, and as a sub-process in the \bpmn\ model in Figure~\ref{fig:bpmn_example}.

{\bf Data}:
\sm{}s and \htn{}s natively support the provision of data inputs to states and tasks. Within \sm{}s, data can be added inside states (both simple and composite), while \htn{}s support the provision of data within the task header or within the set of variables associated with tasks. Concerning \bt{}s, there is no standard way to define parameters for tasks, but some implementations allow adding inputs to action nodes through \textit{ports}.
Instead, \bpmn{}s support \textit{Data Objects}, which represent an object or a collection of objects that can be written and read by the activities in the process. Alternatively, some \bpmn\ implementations allow the definition of the inputs and outputs that are associated with both a single activity and the whole process.
Regarding data storage, the available support is mainly implementation-specific for all the formalisms. Many \bt{} implementations rely on blackboards, a centralized key-value storage, as a mechanism for sharing data between execution nodes. Similarly, \sms{} can leverage global variables, dynamically updated within states. In \bpmn, data can be configured to create process variables that are accessible within the process scope. In contrast, \htn{}s do not provide mechanisms for data storage.

{\bf Communication}:
Among the four formalisms, \bpmn\ is the only one providing support to the explicit modeling of communication in the mission, and specifically through \textit{message} or \textit{signal} events. Message events can represent a one-to-one communication, while signal events can express a broadcast communication~\cite{FaMe}.
In contrast, \bt{}s, \sm{}s, and \htn{}s only rely on the implementation of actions and primitive tasks to realize the communication.

\begin{figure}[t]
    \centering
    \includegraphics[width=0.85\linewidth]{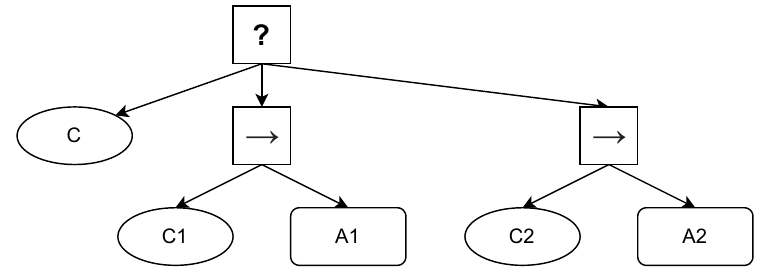}
    \caption{PPA pattern in BTs.}
    \label{fig:ppa}
\end{figure}

{\bf Pre/post-conditions}: \htn{}s natively support their specification both in the tasks and methods definition. A specialization of \sm{}s, namely \textit{Protocol Transition}, enables their support. 
\bpmn\ does not support their specification natively, but some extensions enable it, e.g., in~\cite{IntrigilaPD21,TsaiLW07}. Alternatively, \bpmn intermediate events can be employed as a workaround to constrain task execution based on the satisfaction of certain conditions before or after an activity, although this does not formally capture the semantics of pre- and post-conditions.
Finally, within \bt{}s, \textit{condition nodes} can be used to specify both pre- and post-conditions through the Postcondition-Precondition-Action (PPA) pattern~\cite{colledanchise2018behavior}. Figure~\ref{fig:ppa} shows a general case of PPA: the post-condition $C$ is specified as \textit{condition} node placed as the first child of a \textit{fallback} node, whereas possible actions to reach $C$ are specified within sibling nodes; pre-conditions are specified as \textit{condition} nodes placed as the left sibling of an \textit{action} node with a \textit{sequence} node as a parent (either actions $A1$ or $A2$ can be executed to reach $C$, with pre-conditions $C1$ or $C2$, respectively).

{\bf Events}: 
Even if not explicitly modeled in \bt{}s, their intrinsic reactive nature allows the event handling without any dedicated constructs. In fact, since the tree is continuously ticked, the condition nodes that check for the occurrence of a given event are continuously ticked as well: if a condition node succeeds because of the occurred event, the behavior the robot should exhibit in response can be performed as a consequence. This behavior can be modeled by leveraging the conditional structure shown in Table~\ref{tab:control-flow-op}~\cite{colledanchise2018behavior}. Similarly, in \sm{}s the representation of events is supported by default, since events are the triggers for state changes. Each transition must be associated with an event: if the current active state has an outgoing transition related to the occurred event, the system moves to the target state for handling. This implies that if a recurrent event needs to be handled by different states of the \sm, each state must have an outgoing transition associated with this event. In contrast, \bpmn\ provides elements to explicitly model events that occur within the system. These events can vary in type, such as time-driven, condition-driven, or communication-driven, and can be placed on the boundary of activities, used as starting points for processes or subprocesses, or integrated into the execution flow. Events can also be modeled as interrupting or non-interrupting, meaning that when the event occurs, the main flow is either interrupted or allowed to continue running, respectively.
Finally, \htn{}s do not offer mechanisms for explicitly modeling events, which need to be realized within the robot's mission execution platform.

{\bf Errors}: As a specific case of events, \bts and \sms models errors by leveraging their reactive and event-based nature. \bt handle errors arising from action nodes through the possibly returned \textit{failure} result. Similarly to events, \sms require a transition related to an error from each of the states to properly handle the event during mission execution. \bpmn\ offer \textit{error events} elements for their explicit representation, while \htns do not offer support for error specification at modeling time.

\begin{obs}{\htn event and error handling}\label{obs:htn-reactivity}
While \htns do not provide explicit first-class constructs for events or errors, these aspects can be handled through method preconditions and replanning mechanisms. Events and errors can be addressed at the planning and execution monitoring level, rather than being explicitly modeled in the \htn.
\end{obs}

\subsection{Validation}

To validate the findings related to RQ1, we asked experts to evaluate the correctness and completeness of our comparison for each formalism through two main validation questions (VQ):

\begin{itemize}[labelwidth=3.2em, labelsep=0.6em, leftmargin=!, align=left]
  \item[VQ1.1] \textit{Do you agree with the usage of control structures?}
  \item[VQ1.2] \textit{Do you agree with the modeling of concepts?}
\end{itemize}

For each VQ, respondents rated their agreement on a 5-point Likert scale (1 = strongly disagree, 5 = strongly agree). For responses rated $\leq 3$, participants were required to provide qualitative feedback suggesting clarifications or corrections.
Specifically, for VQ1.1, participants were invited to indicate whether they would suggest modifications or further clarifications. For VQ1.2, they were asked to specify whether any concepts were represented incorrectly or misleadingly, and whether any relevant concepts were missing from the comparison.

\textbf{Results and discussion}: 
The overall agreement scores for BTs, SMs, HTNs, and BPMN are illustrated in \Cref{fig:likert-bts-rq1}, \Cref{fig:likert-sm-rq1}, \Cref{fig:likert-htn-rq1}, and \Cref{fig:likert-bpmn-rq1}, respectively.
Participants mostly agreed with the proposed mapping, with very few disagreements (never more than 3, with at most only one ``strong disagree'' per each formalism-related question).

In the following, we first report broader (minor) concerns that affect all four formalisms, then we discuss the updates for each formalism individually.
Regarding control structures, participants mostly suggested minor refinements to improve the clarity of the descriptions in \Cref{tab:control-flow-op}. For instance, some participants pointed out that the definition of parallel task execution does not coincide with parallel execution in a program. This led us to include Observation~\ref{obs:exec-par} in the text. Moreover, the description of the conditional flow and the mutual exclusivity of guards emerged as a cross-cutting concern for BTs, SMs, and BPMN. We introduced Observation~\ref{obs:mutually-ex} after interviews with \textit{[par:20]}, ~\textit{[par:17]}, ~\textit{[par:10]} ~to clarify this concern. Finally, the explanations of loop modeling in SMs and BPMN were revised to reduce ambiguities and better reflect their realization. These clarifications address some of the lower agreement scores observed for SM and BPMN control structures (see \Cref{fig:likert-sm-rq1} and \Cref{fig:likert-bpmn-rq1}).

For \bts, participants mostly agreed with the proposed mapping, while disagreements (3 for control and 2 for concept modeling, over 22 respondents) were reported for the representation of the conditional control structure (that, initially, did not include checking the $C$ condition and its negation), and for the description of both the parallel node and the loop structures. Several participants highlighted that the initial BT excerpt we provided did not clearly express an \textit{if-else} semantics comparable to that of the other formalisms, and that the conditional constructs across the formalisms are not strictly equivalent due to the semantics of \textit{failure} state. Following the interview with \textit{[par:5]}, we discussed the practical implications of such semantics and refined the modeling and description of the BT conditional control structure to align with the if-else mechanism, leading to Observation~\ref{obs:bt-fail}. Additionally, we discussed best practices for expressing conditions, which we incorporated into the text.

For \sm, only two disagreements were reported by participants (a strong disagree and a disagree for both control structures and concepts, respectively, over 27 respondents). Concerning control structures, they reported errors in the description of the mapping with loop and conditional structures in \Cref{tab:control-flow-op}, which we fixed accordingly. Moreover, some participants questioned the selection policy when multiple condition guards evaluated to true, and the handling of non-satisfied guards (i.e., when none evaluated to true).
As also confirmed by the interaction with \textit{[par:20]}, we observed that the formalism does not enforce mutual exclusivity of guards and default behaviors, leading to Observation~\ref{obs:mutually-ex}.
Concerning the mapping of concepts, it emerged that the mapping of skills to states was debated among respondents. While most agreed on the mapping, others (e.g., \textit{[par:10]}) saw a better-fitting mapping of skills to the state's activity. The following interview with \textit{[par:10]} ~allowed us to clarify this mapping, i.e., skills are represented through states and implemented through \textit{do\_activity} and \textit{entry}/\textit{exit} actions within states. The interview with \textit{[par:5]}\ confirmed that the proposed mapping of tasks and skills is consistent with common practice in robotic \sm{}–based controllers.

For \htn, 3 disagreements were reported overall over the 10 respondents. Besides discussing the possible methods preconditions overlapping (included in Observation~\ref{obs:mutually-ex}), respondents pointed out that event and error handling are handled in practice through preconditions, implementation-based failure handling, and replanning.
We included this in Observation~\ref{obs:htn-reactivity}.

Finally, for \bpmn, besides the discussion on the exclusivity of outgoing conditions in XOR gateways summarized in Observation~\ref{obs:mutually-ex}, the only disagreement (over 14 respondents) case concerned an imprecise description of the loop structure, which has been fixed.
Following the interview with \textit{[par:17]}, we further discussed the possibility of modeling loops through multi-instance tasks. We integrated this alternative representation into \Cref{tab:control-flow-op} and added the corresponding description in the text.
Concerning the concept mapping, a participant raised concerns regarding the expression of pre- and post-conditions in \bpmn. This led us to clarify that, although not natively supported, intermediate events can be used as a workaround to approximate the intended semantics of pre- and post-conditions.

\begin{observation}
\textbf{Summary of RQ1}.
All four formalisms are capable of expressing the fundamental control structures (sequential, conditional, loop, and parallel) through different modeling mechanisms. They differ in how mission concepts are modeled.
\bt and \sm provide explicit constructs for skills and tasks, but offer limited native support for communication or pre- and post-conditions. \htn provides elements for expressing tasks and skills and encodes pre- and post-conditions directly, though data flow and communication remain implicit. \bpmn offers the most explicit and comprehensive support for mission concepts, with dedicated elements for data, communication, events, errors, and task types. However, it does not natively support pre- and post-conditions.
Overall, as highlighted by the experts, all the concepts can be represented either using appropriate modeling patterns or at the implementation level.
\end{observation}

\LikertChart{22}%
  {(0,\CS) (0,\MC)}
  {(3,\CS) (2,\MC)}
  {(3,\CS) (1,\MC)}
  {(11,\CS) (12,\MC)}
  {(5,\CS) (7,\MC)}
  {Likert responses for BTs}
  {fig:likert-bts-rq1}

\LikertChart{27}%
  {(1,\CS) (1,\MC)}
  {(1,\CS) (1,\MC)}
  {(4,\CS) (3,\MC)}
  {(13,\CS) (16,\MC)}
  {(8,\CS) (6,\MC)}
  {Likert responses for SMs}
  {fig:likert-sm-rq1}
  
\LikertChart{10}%
  {(0,\CS) (0,\MC)}
  {(2,\CS) (1,\MC)}
  {(2,\CS) (1,\MC)}
  {(3,\CS) (5,\MC)}
  {(3,\CS) (3,\MC)}
  {Likert responses for HTNs}
  {fig:likert-htn-rq1}

\LikertChart{14}%
  {(1,\CS) (0,\MC)}
  {(0,\CS) (0,\MC)}
  {(1,\CS) (5,\MC)}
  {(6,\CS) (4,\MC)}
  {(6,\CS) (5,\MC)}
  {Likert responses for BPMN}
  {fig:likert-bpmn-rq1}

\section{Peculiarities and Limitations of modeling missions with the formalisms (RQ2)} \label{sec:expressivity}

This section addresses \textit{RQ2} by analyzing the expressiveness of the formalisms concerning how they support the modeling of the concerns related to robotic missions, obtained, as presented in \Cref{sub:methodology-analysis}, by leveraging the scenarios presented in \Cref{sub:methodology-sources}. We first present the identified concerns and discuss how they are supported by the formalisms. Then, we discuss their strengths and weaknesses in modeling robotic missions.

\subsection{Modeling of mission concerns}

From the selected scenarios, we identified the following set of concerns that affect the modeling of robotic missions, stressing the model expressiveness.
Specifically:
\begin{itemize}
    \item \textit{reactive behavior}: behaviors that allow the robot to respond to events or errors requiring the performance of additional actions;
    \item \textit{decision making}: choices made at runtime based on the current system state or overall context;
    \item \textit{time-dependent behavior}: behaviors that have to be executed periodically after a specified interval, triggered after a certain delay, or constrained by timeouts;
    \item \textit{task status}: tracking the execution status of an action,  i.e., if it is completed, ongoing, or if an error occurred;
    \item \textit{robot-robot interaction}: direct interactions between multiple robots involved in the mission, such as inter-robot communication and synchronization;
    \item \textit{human-robot interaction}: direct interaction between humans and robots, involving explicit communication from the robot to the human and vice-versa, e.g., prompting commands/instructions and getting human feedback, or tasks to be executed together with or only by humans;
    \item \textit{robot-external systems interaction}: explicit communication between the robot and external systems, e.g., user interfaces, web services, and databases, for sending/receiving data, commands, etc;
    \item \textit{state saving and task resuming}: pausing and resuming the current mission execution, for allowing its temporary interruption for executing extraordinary actions (e.g., if an event or an error occurs), and later resuming the mission from the interrupted task rather than restarting it from the beginning (also mentioned as \textit{event handler} in~\cite{Promise});
    \item \textit{explicit waiting}: holding the robot in a busy form of waiting for specific events or conditions before starting or proceeding with the execution of the mission;
\end{itemize}

\Cref{tab:scenarios-concerns} reports the concerns identified within each of the scenarios.

\begin{table*}[h]
\centering
\caption{Robotic scenarios and associated concerns. \\
Acronyms used in the table: 
PB (Pick Ball), HR (Humanoid Robot), VS (Vital Signs Monitoring), KC (Keeping Clean), 
FL (Food Logistics), LSL (Lab Samples Logistics), WPH (Welcome People to Hospital), 
DG (Deliver Goods), SUAVE, SA (Smart Agriculture), WA (Warehouse Automation).}
\label{tab:scenarios-concerns}
\renewcommand{\arraystretch}{1.2}
\resizebox{\textwidth}{!}{
\begin{tabular}{l|c|c|c|c|c|c|c|c|c|c|c}
\hline
\textbf{Concern} &
\textbf{PB} &
\textbf{HR} &
\textbf{VS} &
\textbf{KC} &
\textbf{FL} &
\textbf{LSL} &
\textbf{WPH} &
\textbf{DG} &
\textbf{SUAVE} &
\textbf{SA} &
\textbf{WA} \\
\hline
 Reactive behavior & \checkmark & \checkmark & \checkmark & \checkmark & \checkmark &           & \checkmark & \checkmark & \checkmark & \checkmark & \checkmark \\
Decision making   &  & \checkmark  & \checkmark & \checkmark & \checkmark & \checkmark & \checkmark & \checkmark & \checkmark & \checkmark & \checkmark \\
Time-dependent behavior    & &        & \checkmark & \checkmark &            &            &            &            &            &    &     \\
Task status             & &           &            &            &            &            &            & \checkmark &            &        &    \\
Human-robot interaction      & &   \checkmark   & \checkmark &            & \checkmark &           \checkmark & \checkmark &            &            &      & \checkmark    \\
Robot-robot interaction        & &    &            &            & \checkmark &            &            & \checkmark &            & \checkmark &  \\
Robot-external system interaction &    & &         & \checkmark & \checkmark & \checkmark & \checkmark & \checkmark &            &       &  \checkmark \\
State saving \& task resuming     &       & &      & \checkmark &            &            &            & \checkmark & \checkmark &       &   \\
Explicit waiting                  &      & &       &            &            &           \checkmark & \checkmark &            &            &       \checkmark   & \\
\bottomrule
\end{tabular}}
\end{table*}

Each concern is evaluated according to the extent it is supported by each of the formalisms, based on the insights arising from the models obtained from the identified scenario described in \Cref{sec:methodology}. We score the support provided by each formalism for a given concern on three different levels, as follows:

\begin{itemize}
    \item \textit{Full support}, if the formalism provides either native constructs (i.e., elements, operators, or control structures) associated with the concern, or modeling patterns to express it, allowing its modeling to be unambiguous and consistent across the missions without requiring workarounds;
    \item \textit{Partial support}, if the formalism does not provide an explicit or dedicated construct to express the concern, but it can still be modeled indirectly through workarounds or ad hoc solutions that leverage other constructs;
    \item \textit{No support}, if the formalism can not express the concern, neither directly nor through workarounds, hence requiring the realization of such concern using external mechanisms or by realizing it at a different abstraction level. This does not mean that the concern is not addressable when using a given formalism, but that it requires implementation-level effort.
\end{itemize}

\begin{table*}[t]
\renewcommand{\arraystretch}{1.5}
\centering
\caption{Summary of formalism expressivity for mission concerns. \\ (\full: full support; \half: partial support; \emptycirc: no support)}
\begin{tabular}{p{2.2cm}ccccp{11.1cm}}
\toprule
\textbf{Mission concern} & \textbf{BT} & \textbf{SM} & \textbf{HTN} & \textbf{BPMN} & \textbf{\textit{Rationale}} \\
\midrule
\textbf{Reactive behavior} & \full & \full & \emptycirc & \full & \bt and \sm leverage their reactive nature and event-handling structures. \bpmn feature boundary events and event sub-processes. \htn does not support it.\\
\textbf{Decision making} & \full & \full & \full & \full & \sm and \bpmn feature conditional control structures for runtime decision-making. In \bts it is realized by combining fallbacks and sequence nodes. \htn relies on planning according to pre- and post-conditions associated with methods. \\
\textbf{Time-dependent behavior} & \half & \half & \emptycirc & \full & In \bt and \sm it has to be realized by manually implementing the control of timing through condition nodes and state implementations, respectively. \bpmn features timer events. \htn does not support it. \\
\textbf{Task status} & \full & \full & \emptycirc & \full & \bt, \sm, and \bpmn support it through the value returned by the tick, the events outgoing from a state, and the activity lifecycle, respectively. \htn does not support it.\\
\textbf{Robot-robot interaction} & \half & \half & \half & \full & \bt, \sm, and \htn rely on the action nodes, states, and tasks implementation, respectively. \bpmn has dedicated structures, i.e., message and signal events for communication.\\
\textbf{Human-robot interaction} & \half & \half & \half & \full & \bt, \sm, and \htn rely on the action nodes, states, and tasks implementation, respectively. \bpmn supports user and manual tasks, and the explicit modeling of communication with message/signal events. \\
\textbf{Robot-external system interaction} & \half & \half & \half & \full & \bt, \sm, and \htn rely on the action nodes, states, and tasks implementation, respectively. \bpmn has dedicated task types (service task, send task, receive task) and message/signal events. \\
\textbf{State saving and task resuming} & \full & \half & \emptycirc & \emptycirc & \bt offers control flow nodes with memory to keep track of the overall status of the mission, or can rely on a shared knowledge (blackboards) to keep track of already executed tasks. \sm requires ad hoc states to serve as history states and events to first pause and then resume the execution. \htn and \bpmn do not support it. \\
\textbf{Explicit waiting} & \half & \half & \half & \full & \bt, \sm, and \htn rely on the ad hoc implementation of action nodes, states, and tasks, respectively, since they do not provide dedicated elements. \bpmn supports intermediate events of different types. \\
\bottomrule
\end{tabular}
\label{tab:expressivity}
\end{table*}

\Cref{tab:expressivity} summarizes the support provided by each of the formalisms in expressing the mission concerns.

{\bf Reactive behavior}: 
Regarding the expression of the reactive behavior, \bts, \sms, and \bpmn fully support its modeling, as a direct effect of their reactive nature (\bt{}s and \sm{}s) or by leveraging event elements (\bpmn).
In particular, \bts support this through the continuous tree ticking, which allows the evaluation of the tree and its associated condition nodes at each tick, hence enabling the execution of the actions guarded by condition nodes that check the occurrence of a certain event, by following the conditional control structure in \Cref{tab:control-flow-op}.
\begin{figure}[h!]
    \centering
    \includegraphics[width=0.6\linewidth]{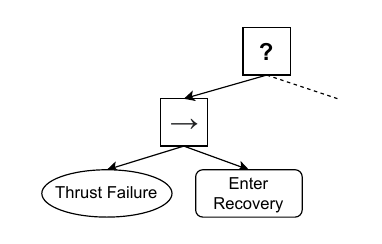}
    \vspace{-1em}
    \caption{Example of reactive behavior realized through \bt.}
    \label{fig:reactive-bt}
\end{figure}
\Cref{fig:reactive-bt} shows an excerpt of the \bt modeling the \textit{SUAVE} mission, where the (mission-wide) reactive behavior is controlled through a \textit{fallback} operator placed as a root of the subtree modeling the tasks that have to be preempted when an event occurs, and a sequence node as child (placed left-most to the tasks to be preempted): when the condition node's tick returns \texttt{Success}, the associated response behavior is executed (i.e., when \texttt{Thrust Failure} holds, then \texttt{Enter Recovery} is executed). Otherwise, the tick is propagated to the next child of the \textit{fallback} node.
On their side, \sms support this through the intrinsic event-based nature of the model, where events drive the transitions between states. In this case, transitions (labeled with the event to react to) connect the system states to be preempted to the state(s) modeling the actions to perform as a response.
In \bpmn, the reactive behavior is expressed by exploiting boundary events or event sub-processes. Boundary events are attached to activities that may require a reaction to events and enable the execution flow to directly transition to other activities, modeling the corresponding response. Event sub-processes can be employed to handle events or errors that may occur at any point during the mission execution. Both strategies allow actions in response to events to be executed either interrupting the ``normal'' mission execution flow or as a parallel process.
Conversely, \htn does not provide support for reactive behavior during the runtime.

{\bf Decision making}:
Concerning the modeling of the runtime decision-making of the system, all the formalisms, although at different levels, support it by applying the \textit{conditional} control structures reported in \Cref{tab:control-flow-op}.
In particular, \bt, \sm, and \bpmn have explicit control structures to control the runtime behavior by switching between different tasks according to the runtime conditions. \htn, on the other side, relies on the pre-conditions of the defined methods to define alternative behaviors. The association of methods to abstract tasks is done through planning~\cite{ghallab2016automated}, which needs to be performed at runtime in order to consider runtime conditions that are not accessible beforehand.
\begin{figure}[t!]
    \centering
    \includegraphics[width=\linewidth]{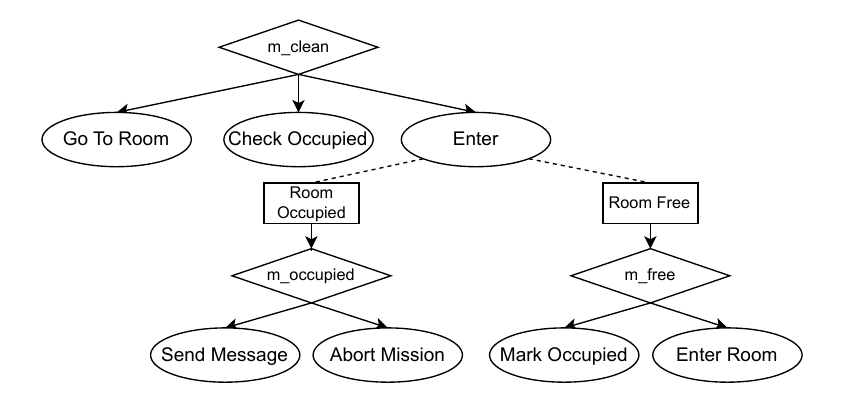}
    \vspace{-1em}
    \caption{Example of decision-making realized through \htn.}
    \label{fig:decision_making-htn}
\end{figure}
\Cref{fig:decision_making-htn} shows an excerpt of the \htn modeling the \textit{Keeping Clean} mission, where the robot's actions refining the abstract task \texttt{Enter} have to be decided according to the room's status. In this case, two different methods are defined, associated with different pre-conditions: \texttt{m\_occupied} defining the behavior when the room is occupied, \texttt{m\_free} when the room is free. Runtime planning takes into account such pre-conditions to associate the proper method to refine the abstract task \texttt{Enter}.

\begin{figure}[t!]
    \centering
    \includegraphics[width=0.6\linewidth]{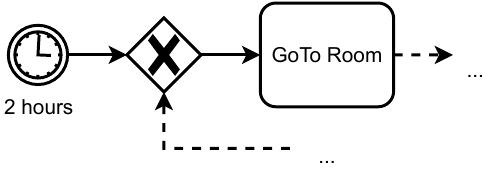}
    \caption{Example of time-dependent behavior realized through \bpmn.}
    \label{fig:timer-bpmn}
\end{figure}

{\bf Time-dependent behavior}:
Concerning the modeling of time-dependent behavior like timeouts and time triggers, \bts and \sms have to rely on specific implementations of action and conditions nodes (\bts), or states and events (\sms) that check ad hoc realized timers and react to them consequently, as described for the \textit{reactive behavior}.
\htn does not support this feature. \bpmn provides timer events, which can be defined either for a specific date and time or for a duration (e.g., every two hours). These events can be used in different parts of the mission to constrain the start of the process to a given time, act as interrupting triggers during execution, or pause the flow for a specified duration.
\Cref{fig:timer-bpmn} shows the initial part of the \textit{Vital Signs Monitoring} scenario, which prescribes that all patients’ vital signs be checked every two hours. In \bpmn, this periodicity is captured using a timer start event, which triggers the mission execution at the required two-hour interval.

\begin{figure}[b!]
    \centering
    \includegraphics[width=0.9\linewidth]{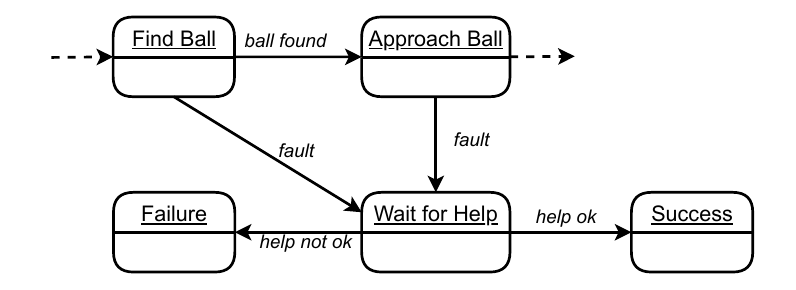}
    \caption{Example of task status handling with \sm.}
    \label{fig:taskstatus-sm}
\end{figure}

{\bf Task status}:
Concerning the ability of keeping track of the status of a mission task, \bts support this through the returned value of the tick on a node: \textit{Success}, \textit{Running}, or \textit{Failure}, as described in~\Cref{sec:background}. Interestingly, it is worth noting that the \textit{Failure} value returned by condition nodes has a different semantics than the one returned by action nodes: the first indicates a condition that is currently not holding; the second indicates possible failures during the action execution.
\sms do not prescribe predefined execution statuses; however, task status can be represented either by dedicated states (e.g., \textit{Success}, \textit{Failure}) or by the events or outcomes emitted by a state, which may trigger transitions to different successor states depending on whether a task completes successfully or fails. \Cref{fig:taskstatus-sm} shows an excerpt of the \sm modeling the \textit{Pick Ball} scenario in which task failure is modeled both via outgoing transitions labeled as \texttt{fault} and through dedicated states for the recovery (\texttt{Wait for Help}). In this case, if a fault occurs within the \texttt{Find Ball} or \texttt{Approach Ball} states, the robot waits for help. Afterwards, if the help was successful (\texttt{help ok} transition) the robot goes in the \texttt{Success} state; otherwise, it goes in the \texttt{Failure} state.

In \bpmn, task status is encoded in the activity lifecycle, which includes states such as \textit{Ready}, \textit{Active}, \textit{Withdrawn}, \textit{Completed}, and \textit{Failed}. Transitions between these states determine how tokens progress through the process and enable the specification of different behaviors depending on the execution outcome of an activity~\cite[p.428]{omg_business_2011}.
Differently, \htn does not handle this concern.

\begin{figure}[t!]
    \centering
    \includegraphics[width=0.55\linewidth]{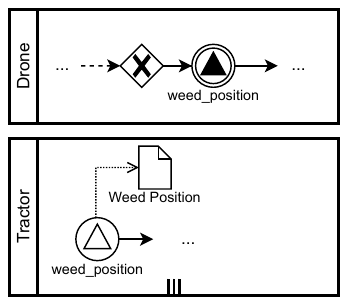}
    \caption{Example of robot-robot interaction with \bpmn.}
    \label{fig:r2r-bpmn}
\end{figure}

{\bf Robot-robot interaction}:
The interaction with other robots is not explicitly supported by \bts, \sms, and \htns, as they do not offer ad hoc constructs to model the communication with other parties. To realize this kind of interaction, such formalisms must rely on the implementation of the action nodes (\bt), states (\sm), and tasks (\htn) which have to be realized ad hoc. Conversely, \bpmn offers different constructs allowing interactions, such as explicit message events for one-to-one communication, or signal events, which can be exploited for explicitly modeling multicast communication among multiple robots. \Cref{fig:r2r-bpmn} shows an excerpt of the model built for the \textit{Smart Agriculture} scenario, where the robot-to-robot communication is modeled through a signal event: the \texttt{Drone} shares the position of a weed grass via the \texttt{weed\_position} signal send event, and a currently active \texttt{Tractor} can catch it through the signal receive event.

\begin{figure}[b!]
    \centering
    \includegraphics[width=0.8\linewidth]{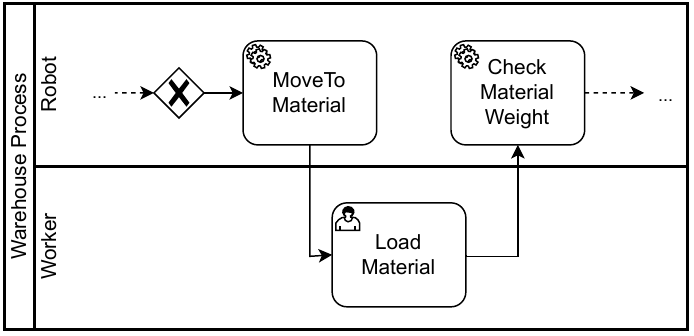}
    \caption{Example of robot-human interaction with \bpmn.}
    \label{fig:r2h-bpmn}
\end{figure}

{\bf Human-robot interaction}: 
Similarly to the previous concern, \bts, \sms, and \htns, have to rely on the implementation of action nodes, states, and tasks, respectively, to explicitly model this kind of concern. \bpmn\ offers a richer set of modeling elements that enable explicit representation of human involvement. \textit{User} and \textit{Manual tasks}~\cite[p.160]{omg_business_2011} allow the specification of activities performed by humans, either with system involvement in the case of User Tasks, or without system support in the case of Manual Tasks. Moreover, human behavior can be integrated directly into the mission model by assigning it to a dedicated lane that executes tasks interleaved within the robot’s workflow~\cite{ReyCCMC19}(see \Cref{fig:r2h-bpmn}, showing an excerpt of the \textit{Warehouse} scenario), or by defining a separate interacting process that exchanges messages with the robot through message or signal events.

{\bf Robot-external system interaction}: 
Similarly, \bts, \sms, and \htns have to rely on the implementation of the action nodes, states, and tasks to model this kind of concern. 
\bpmn offers dedicated mechanisms that allow these interactions to be modeled directly within the process. The approach is analogous to the one used for human involvement, with the main difference lying in the specific task types employed. 
Interactions with external systems can be modeled using \textit{Service Tasks}, which represent automated operations carried out by external software components, or through message flows that capture communication between the robot process and external participants or system components. Notably, Service Tasks may also be used to invoke a corresponding robotic activity, such as navigation, via connectors when the \bpmn process executes outside the robot itself~\cite{ReyCCMC19,de2020event}.

\begin{figure}[t!]
    \centering

    \subfloat[Time-based waiting.\label{fig:explicit_waiting_time-bt}]{
        \includegraphics[width=0.6\linewidth]{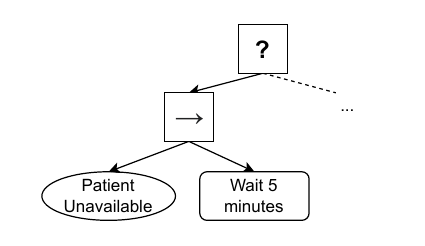}
    }

    \vspace{1em}

    \subfloat[Event-based waiting.\label{fig:explicit_waiting_event-bt}]{
        \includegraphics[width=\linewidth]{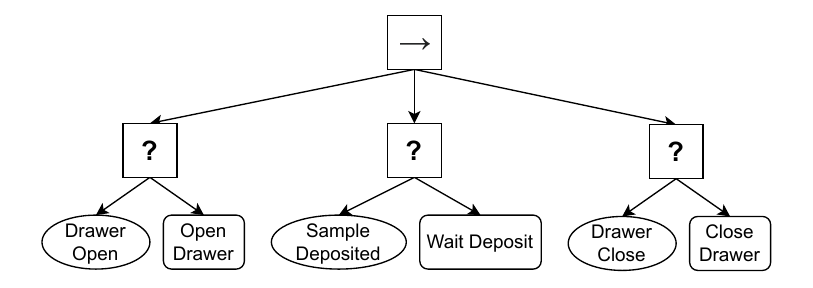}
    }
    \caption{Examples of explicit waiting realized through \bt.}
    \label{fig:explicit_waiting-bt}
\end{figure}

\begin{table*}[t!]
\centering
\caption{Overview of formalism expressivity strengths and weaknesses.}
\begin{tabular}{p{0.4\textwidth}|p{0.54\textwidth}}
\toprule
\multicolumn{1}{c}{\textbf{Strengths}} & \multicolumn{1}{c}{\textbf{Weaknesses}} \\
\midrule 
\multicolumn{2}{c}{\textbf{\bt}}\\
\midrule
$\bullet$ Easily express reactive behavior when the robot has to continuously react to changing conditions.\newline
$\bullet$ Clear task status semantics, making progress and failure handling a first-class concern.
&
$\bullet$ The semantics \textit{Failure} is overloaded, making action and condition nodes returning \textit{Failure} for different reasons (system-level failures or conditions not holding), requiring additional disambiguation through ad hoc nodes to catch possible errors.\newline
$\bullet$ No standard way for expressing waitings, temporal constraints, and interactions, which must be manually implemented in the action/condition nodes.\newline
$\bullet$ The use of nodes with memory for keeping track of mission execution state hinders the tree reactivity by limiting the overall tree re-evaluation.
\\
\midrule 
\multicolumn{2}{c}{\textbf{\sm}}\\
\midrule
$\bullet$ Natively model event-driven behavior defining how the system should react to events and transition between states\newline
$\bullet$ Easily handles task status as distinct states and via transitions outgoing from them.
&
$\bullet$ Handling errors and events requires outgoing transitions from all the potentially affected states, hence requiring explicit transitions from every state for handling system-level events.\newline
$\bullet$ Task resuming requires ad hoc history states and transitions towards each state.\newline
$\bullet$ Interactions, messaging, and coordination with other robots, systems, or humans have to be manually realized.\newline
$\bullet$ Time-dependent behaviors require the manual integration of timers within the state implementation.
\\
\midrule 
\multicolumn{2}{c}{\textbf{\htn}}\\
\midrule
$\bullet$ Provides support for decision making through runtime planning, achieved by decomposing tasks into sub-tasks using methods.\newline
$\bullet$ Natively allows explicit definition of pre- and post-conditions.
&
$\bullet$ No expressiveness for reactive behavior, which has to be realized at a different level outside of the mission model.\newline
$\bullet$ No explicit expression of static choices, which are always demanded to planning.\newline
$\bullet$ No task status tracking, which is totally demanded to the \htn executor implementation.\newline
$\bullet$ Interactions, waiting, and time-driven behaviors are not natively supported and must be manually implemented within task definitions.
\\
\midrule 
\multicolumn{2}{c}{\textbf{\bpmn}}\\
\midrule
$\bullet$ Decision-making logic is easy to model thanks to the process-oriented structure.\newline
$\bullet$ Rich notation explicitly supporting diverse event and error handling, making it very expressive for time-dependent behavior, event/error handling, and waiting. \newline
$\bullet$ Explicitly allows modeling of human activities and external systems, and offers a clear representation of robot-robot, human-robot, and robot-system interactions.\newline
&
$\bullet$ No native support for specifying tasks or mission resuming after they are interrupted.\newline
$\bullet$ Dealing with reactions to highly-frequent events can not be optimally achieved. \newline
$\bullet$ Can become complex and overloaded for detailed and extended robotic missions.
\\
\bottomrule
\end{tabular}
\label{tab:pros-cons}
\end{table*}

{\bf State saving and task resuming}: 
State saving and task resuming require the mission to be first paused (state saving) and then resumed from the point where it was paused (task resuming). As explained, this concern allows the execution of exceptional behavior in response to an event, as discussed for the reactive behavior modeling, and the restoration of the normal mission behavior afterward. Interestingly, none of the constructs fully support this concern explicitly.
\bts can achieve the resuming of the task execution by either (i) using control flow nodes with memory~\cite{colledanchise2018behavior} to keep track of the ongoing state of the mission by internally storing the results returned by their children, until the node returns \textit{Success} or \textit{Failure} to its parent, or (ii) designing the whole tree in such a way, before ticking an action node, a condition node checks the overall mission status using the \textit{backchaining} paradigm~\cite{colledanchise2018behavior}. In both cases, the solution avoids nodes from being ticked again if previously completed (i.e., avoids tasks from being re-executed when not needed), hence allowing the mission to be resumed from where it was interrupted. However, both solutions have limitations: in the first case, the use of control flow nodes with memory limits the overall reactivity of the tree~\cite{millington2019ai,colledanchise2018behavior}, while, in the latter case, task status may have to be manually stored into separate structures, such as \textit{blackboards}, which are not part of the formalism, although commonly supported by \bt implementations.
Differently, \sms need ad hoc event handling, with events outgoing from states, for pausing the mission execution (as described for the reactive behavior), while state transitions labelled with the previously interrupted task have to be redirected to the corresponding state for their resuming. This requires that the information about the previously-interrupted task has to be manually carried on~\cite{IovinoFFCSS25}.
In any case, the pause of the actions within the currently running node or the currently active state has to be manually implemented within the event response. However, being the support offered by \bts more advanced than \sms, since \bts provide built-in nodes with memory, we evaluated them as offering \textit{full support}, while \sms as offering \textit{partial support}.
\htn and \bpmn, in contrast, do not offer solutions or workarounds to address this concern.

\begin{obs}{State saving and task resuming in \bpmn}\label{obs:bpmn-task-saving}
State saving and task resuming can be realized at the execution engine level by storing the process-instance state (e.g., tokens position and the status of activity instances) upon interruption, and reinstating it to resume the process from the same point.
Moreover, \bpmn compensation mechanisms could support rollback-like behavior, though their application to robotic missions remains largely unexplored.
\end{obs}

{\bf Explicit waiting}: 
\bt, \sm, and \htn do not offer constructs for modeling an explicit busy-form of waiting during the mission execution. All of them have to rely on ad hoc implementations of action nodes, states, and tasks, respectively. In contrast, \bpmn events can deal with this concern. Indeed, events can be placed within the process flow to model waiting for different situations, such as a specified duration, the satisfaction of a condition, or the receipt of a message or signal event.
\Cref{fig:explicit_waiting-bt} illustrates two \bts modeling an example of time-based waiting from the \textit{Vital Signs Monitoring} scenario. Here, the timer event requires that, after assessing that the patient is not available and leaving the room, the robot wait for 5 minutes before re-entering the room. In this case, the mission is paused until 5 minutes have elapsed.
Here, the time-based waiting (\Cref{fig:explicit_waiting_time-bt}) is realized by leveraging the conditional control structure reported in \Cref{tab:control-flow-op}: if the patient is not available, then the 5-minute waiting is realized through an ad hoc action node whose implementation has to keep the robot in a waiting state by returning \textit{Running} until 5 minutes are elapsed. Differently, the tree enabling event waiting (\Cref{fig:explicit_waiting_event-bt}) can be realized\footnote{We recall that there is not only a single way to realize a given behavior. We report one out of the many possible modeling solutions.} by pairing each action node with a condition node through a \textit{fallback} (\textit{explicit success condition} pattern in~\cite{colledanchise2018behavior}). In particular, after having opened the drawer, the robot checks if the sample is deposited and, if not, the \texttt{Wait Deposit} action realizes the waiting by idling over a \textit{Running} response returned at each tick. By leveraging the reactive nature realized through the continuous tree ticking, after the sample is deposited, the \texttt{Sample Deposited} condition will return \textit{Success} and the execution proceeds towards the next branch (controlling the drawer closing).

\begin{obs}{Implementation-specific concerns} \label{obs:implementation}
    Time-dependent behaviors, explicit waiting, and state saving and task resuming can be realized by leveraging formalism elements in ad hoc solutions. E.g., ad hoc-defined decorators for waiting in \bts, blackboards or nodes with memory and history state nodes for state saving and task resuming in \bts and \sms, respectively.
    Some implementations of \bt, \sm, and \htn provide extensions enabling the modeling of the aforementioned concerns.
\end{obs}

\subsection{Formalism expressivity strengths and weaknesses}

\Cref{tab:pros-cons} reports the main strengths and weaknesses of the formalisms with respect to the expressiveness of the identified concerns. They have been drawn from the analysis of the expressiveness conducted in the previous section.

{\bf \bt}:
In general, \bt's major strength lies in the expression of reactive behavior, thanks to the tick-based execution strategy that allows for the continuous evaluation of the entire tree. This makes the evaluation of mission-level conditions and the monitoring of events (i.e., those conditions that must be checked throughout the whole mission, and the events that affect the overall mission) easy to specify within the model.
Moreover, the \textit{Success}, \textit{Failure}, or \textit{Running} result returned by all the nodes after each tick allows the continuous evaluation and control of the mission execution status. On the other side, as also mentioned in Section~\ref{sec:concepts}, the \textit{Failure} state's semantic results overloaded since it is returned by both action and condition nodes for two possibly different reasons: for action nodes, it is usually associated with failures in the action being executed, while for condition nodes it is associated to a condition that is currently not met (see, for instance, the conditional structure in \Cref{tab:control-flow-op} or the PPA pattern in \Cref{fig:ppa}, leveraging the conditional nodes for controlling the tree ticking). This overloading requires disambiguation to avoid the execution of unintended behavior. For instance, if the \texttt{Enter Recovery} action in \Cref{fig:reactive-bt}, for some reason, fails, the resulting \textit{Failure} can be interpreted by the \textit{fallback} node at the tree root the same way as if there was no thrust failure, causing the mission to proceed as if no failure was detected.
Also, \bt lacks dedicated constructs for temporal-related behavior and interactions with other robots, systems, and humans. They have to be manually implemented within the action nodes.

{\bf \sm}:
Conversely, from \bts, \sms employs transitions from states to explicitly distinguish between different events and different outcomes of the actions executed by the robots within each state. This allows the mission model to unambiguously distinguish different events and error causes, and to explicitly redirect the flow towards the states modeling their handling. As a drawback, handling events that can potentially affect every state requires outgoing transitions from each of them, hence making the model more complex. Organizing states hierarchically~\cite{harel1987statecharts} (as in \Cref{fig:sm_example}) allows reducing such complexity. Similarly, for resuming a temporarily interrupted mission, transitions towards all the possible interrupted states have to be modeled. This also requires the specification of ad hoc \textit{selector} states~\cite{IovinoFFCSS25}, hence contributing to the model complexity.
As for \bts, the handling of temporal-related behaviors has to be manually realized within the single state logic, as well as the interaction with other robots, systems, and humans.

{\bf \htn}: 
\htns have the strength (unique among all the considered formalisms) of explicitly expressing pre- and post-conditions, which provide the support for binding abstract tasks to multiple methods that can realize decision-making through runtime planning. The task decomposition obtained by refining abstract tasks into methods naturally realizes a modular structure. As a drawback, choices can not be embedded explicitly into the model, since planning is always required.
Moreover, no support for reactiveness and no mechanisms to handle task status are provided within the model definition, relegating this feature to both external documentation/modeling and behavior implementations.

{\bf \bpmn}:
\bpmn provides the richest expressivity among the considered formalisms, thanks to its comprehensive standard notation, which enables many mission-relevant concepts to be modeled explicitly.
Communication can be represented through message and signal events, while the variety of task types allows the specification of interactions with humans, robots, and external systems. Decision-making and control logic are naturally expressed through gateways. 
However, \bpmn offers limited support for resuming tasks after interruption: once a token leaves an activity, its execution state is lost, requiring the activity to restart.
As discussed in Observation~\ref{obs:bpmn-task-saving}, a possible workaround can be realized by acting on the process execution engine, by extending it to keep track of the executed tasks. However, this should be implemented on top of BPMN execution engines.
Reacting to frequent events or modeling fine-grained robotic missions often leads to large and complex diagrams. For instance, continuous condition monitoring (e.g., battery level) can be modeled with event subprocesses, but their execution typically overrides the main flow unless the engine provides specialized handling. Alternatively, one may attach boundary events to each task or add further gateways, both of which increase model complexity.

\LikertChartVQTwoOne{29}
{(0,\RB) (0,\DM) (0,\TDB) (0,\TS) (0,\RRI) (0,\HRI) (0,\RESI) (0,\SSTR) (0,\EW)} 
{(1,\RB) (0,\DM) (7,\TDB) (0,\TS) (0,\RRI) (1,\HRI) (2,\RESI) (6,\SSTR) (4,\EW)} 
{(2,\RB) (1,\DM) (2,\TDB) (0,\TS) (3,\RRI) (4,\HRI) (4,\RESI) (2,\SSTR) (3,\EW)} 
{(7,\RB) (10,\DM) (12,\TDB) (13,\TS) (10,\RRI) (9,\HRI) (10,\RESI) (11,\SSTR) (10,\EW)} 
{(19,\RB) (18,\DM) (8,\TDB) (16,\TS) (16,\RRI) (15,\HRI) (13,\RESI) (10,\SSTR) (12,\EW)} 
{Likert responses for VQ2.1}
{fig:likert-vq21}

\LikertChartVQTwoTwo{27}
{(1,\bt) (0,\sm) (0,\htn) (0,\bpmn)} 
{(2,\bt) (1,\sm) (1,\htn) (0,\bpmn)} 
{(1,\bt) (3,\sm) (0,\htn) (2,\bpmn)} 
{(13,\bt) (17,\sm) (6,\htn) (9,\bpmn)} 
{(5,\bt) (6,\sm) (3,\htn) (3,\bpmn)} 
{Likert responses for VQ2.2}
{fig:likert-vq22}

\subsection{Validation}

To validate the findings related to RQ2, we asked experts to evaluate their agreement on the support of formalism expressivity and with the identified strengths and weaknesses. 
Hence, for each identified mission concern, we defined the following VQ:
\begin{itemize}[labelwidth=3.2em, labelsep=0.6em, leftmargin=!, align=left]
    \item [VQ2.1] \textit{Do you agree with the formalism expressivity for the concern?}
    \item [VQ2.2] \textit{Do you agree with the identified strengths and weaknesses?}
\end{itemize}
For each VQ, respondents rated their agreement on a 5-point Likert scale. For responses rated $\leq 3$, participants were required to provide qualitative feedback by explaining the reason for the disagreement.
Specifically, for VQ2.1, participants were asked to rate their agreement with each of the identified mission concerns, while VQ2.2 was asked for each formalism they acknowledged expertise with.

\textbf{Results and discussion}:
\Cref{fig:likert-vq21} and \Cref{fig:likert-vq22} show the overall agreement scores obtained for VQ2.1 and VQ2.2, respectively. Most of the participants agreed with both the expressivity for the mission concerns and for the strengths and weaknesses.

Concerning expressivity (VQ2.1), some disagreements were associated with the time-dependent behavior, the state saving, and the representation of the explicit waiting (7, 6, 4 over 29 respondents, respectively).
In particular, they argued that some implementations of \bt, \sm, and \htn provide extensions to enable the modeling of time-dependent behavior, state saving and task resuming, and explicit waiting, and that such concerns can be realized without difficulty using the formalism elements. We included this in Observation~\ref{obs:implementation}.
Additionally, some respondents noted that, for \bpmn, state saving and task resuming could potentially be implemented at the execution-engine level. The follow-up discussion with \textit{[par:17]} ~confirmed this possibility, and further hinted at the possibility of using \textit{compensation} for having rollback-like behaviors, leading to Observation~\ref{obs:bpmn-task-saving}.
Moreover, a small number of respondents (3 over 29) disagreed on the human-robot and robot-external system interactions. As motivation, they suggested that interactions can be, in general, considered as normal actions and modeled as for other skills. Follow-up discussions with \textit{[par:5]} ~also clarified that interactions with external systems can also be interpreted as part of the interactions with the robot’s environment, where an implicit communication is mediated through sensing actions. Similarly, some respondents argued that explicit waiting can be modeled using standard nodes, delegating the waiting logic to their execution-level implementation.
After follow-up discussions, interviewed participants converged toward agreement with the assigned levels of support for the discussed concerns, as, per the defined support levels, the absence of dedicated explicit structures is marked as ``partial support''.

\begin{table*}[!htbp]
\centering
\caption{Identified tools for \bt (Grey-highlighted rows indicate baseline tools supporting the core formalism and used as foundations for ROS-based packages).}
\label{tab:tools_bt}
\resizebox{\linewidth}{!}{
{\footnotesize
\begin{tabular}{p{0.12\linewidth}p{0.13\linewidth}p{0.39\linewidth}p{0.26\linewidth}}
\toprule
\textbf{Name} & \textbf{Scope} & \textbf{Description} & \textbf{Documentation Reference} \\
\midrule \midrule
\rowcolor{mygray} BehaviorTree.CPP & Execution; Debugging & C++ library for executing \bt. Provides a rich set of control nodes, logging facilities, and interfaces for debugging purposes.  & \exLink{behaviortree.dev}{https://www.behaviortree.dev}\\ \hline
\rowcolor{mygray} PyTrees & Execution; Monitoring; Debugging & Python library for executing \bt, designed to facilitate the rapid development of medium-sized decision-making engines. Offers minimal visualizations for monitoring and debugging purposes. & \exLink{py-trees.readthedocs.io}{https://py-trees.readthedocs.io/en/devel/} \\ \hline
\rowcolor{mygray} Groot & Modeling; Monitoring; Debugging & Graphical editor for \bt. Allows tree design, log playback, and real-time introspection when connected to BehaviorTree.CPP. & \exLink{behaviortree.dev/groot}{https://www.behaviortree.dev/groot} \\ \hline
\rowcolor{mygray} Forester & Execution & Orchestration engine implementing \bt. Supports exporting trees for use with the ROS navigation library. & \exLink{forester-bt.github.io}{https://forester-bt.github.io/forester/} \\ \hline
BehaviorTree.ROS2 & Execution & ROS-compatible implementation of BehaviorTree.CPP & \exLink{behaviortree.dev/docs/ros2\_integration}{https://www.behaviortree.dev/docs/ros2_integration/}\\ \hline
PyTrees ROS & Execution & ROS-compatible implementation of PyTrees. & \exLink{py-trees-ros.readthedocs.io}{https://py-trees-ros.readthedocs.io/en/devel/} \\ \hline
ros2\_ros\_bt\_py & Execution; Monitoring &
Python library supporting runtime execution and integration with ROS2 components, with introspection capabilities for monitoring behavior. &
\exLink{fzi-forschungszentrum-informatik.github.io/ros2\_ros\_bt\_py/}{https://fzi-forschungszentrum-informatik.github.io/ros2_ros_bt_py/} \\
\bottomrule
\end{tabular}
}}
\end{table*}

Concerning the identified strengths and weaknesses (VQ2.2), respondents were mostly concerned about the ones identified for \bts (1 strong disagree, 2 disagree), mostly arising from the following (now removed) weakness: ``it is difficult to trace the origin of failures returned by nodes deeper in the tree''.
After follow-up interactions with \textit{[par:5]}, we removed the statement from \Cref{tab:pros-cons} because it applies to all the considered formalisms, particularly those supporting hierarchy, and more generally to programs using \textit{try-catch} mechanisms. Moreover, the first weakness listed for \bts\ in \Cref{tab:pros-cons} already captures this aspect through the use of dedicated nodes for error handling.
Regarding \sms, the only reported disagreement concerned the treatment of task resuming.
After follow-up discussions with \textit{[par:10]}\ and \textit{[par:20]}, we clarified that history states can be used to model task saving and resuming, and we reflected this both in the text and in \Cref{tab:pros-cons}.

\begin{observation}
\textbf{Summary of RQ2}.
The four formalisms have different expressivity properties. Overall, \bt and \sm, even though not always explicitly, are able to cover all the considered mission concerns by leveraging their base elements. \htn resulted as the more ``rigid'', being unable to express reactive behavior, and delegating to runtime planning the decision-making capabilities. \bpmn offers the richest set of elements, although the models can be complex if the mission specification is fine-grained.
The identified concerns, even if not explicitly supported, could be addressed either by relying on workarounds leveraging the existing formalism elements (for partially-supported concerns), or at the implementation level through execution-layer mechanisms (for non-supported concerns).

\end{observation}

\section{Available tools supporting the formalisms (RQ3)}\label{sec:tools}

This section provides the analysis and evaluation of tools supporting the four formalisms. We first analyze the landscape of tools supporting the formalisms, focusing on those that are actively maintained and on their scope and usability in robotic missions.

\subsection{Tools analysis}

Tables~\ref{tab:tools_bt}, \ref{tab:tools_sm}, \ref{tab:tools_htn} and \ref{tab:tools_bpmn} report the identified tools for each formalism, together with their scope, a brief description, and the corresponding documentation references. The rows highlighted in grey denote the tools that, although not designed and realized for robotic systems, have served as the foundation for the development of current ROS-compatible packages.
Regarding the scope, we distinguish five objectives that a tool may support: modeling, execution, monitoring, debugging, and planning. \textit{Modeling} indicates that the tool provides graphical interfaces to create, edit, or visualize mission specifications using the constructs of the corresponding formalism. \textit{Execution} refers to the tool’s ability to interpret, execute, or run the specified model, either as a standalone engine or as a component integrated within ROS.
\textit{Monitoring} refers to the runtime information during execution, offering real-time insights into mission status and ongoing tasks. \textit{Debugging} supports developers in diagnosing undesired behaviors through features such as trace visualization, state status, and logging of internal execution events. In addition to these scopes, tools for \htn{}s may also support a dedicated \textit{planning} capability. This refers to the automatic generation of a task decomposition or action sequence from an \htn\ domain description. 

{\bf BT}: is supported by several tools (see Table~\ref{tab:tools_bt}). Among the most mature and used ones, \textit{BehaviorTree.CPP}, \textit{PyTrees}, and \textit{Groot} provide graphical editors, execution engines, logging tools, and visualization support that facilitate the design and runtime analysis of \bt-based behaviors. Moreover, ROS integration is natively supported through dedicated packages, i.e., \textit{BehaviorTree.ROS2} and \textit{PyTrees ROS}.
Similarly, the \textit{Forester} tool provides a \bt\ engine that enables the definition of behavior trees through its own DSL. Although it does not include ROS-specific components, it supports exporting a Forester tree into a format compatible with the ROS navigation library.
Additionally, \textit{ros2\_ros\_bt\_py} is a ROS2-based Python library for defining and executing BTs. The library enables the specification of BTs directly in code and provides tight integration with ROS2 components, supporting execution and runtime monitoring.
In addition to these tools, other solutions, such as CoSTAR~\cite{CoSTAR}, have been developed as prototypes released together with academic publications. As a consequence, they do not provide long-term maintenance or general-purpose applicability.
Finally, it is worth noting the \exLink{github.com/narcispr/py\_trees\_meet\_groot}{https://github.com/narcispr/py_trees_meet_groot} module that enables loading Groot-generated BTs into the PyTrees library, thus providing an automatic mapping from BehaviorTree.CPP semantics to PyTrees one.

\begin{table*}[!htbp]
\centering
\caption{Identified tools for \sm.
}
\label{tab:tools_sm}
\resizebox{\linewidth}{!}{
{\footnotesize
\begin{tabular}{p{0.12\linewidth}p{0.13\linewidth}p{0.39\linewidth}p{0.26\linewidth}}
\toprule
\textbf{Name} & \textbf{Scope} & \textbf{Description} & \textbf{Documentation Reference} \\
\midrule \midrule
SMACH & Execution; Monitoring & Python library for implementing hierarchical \sm\ in ROS. Provides a runtime viewer that displays active states but offers no modeling interfaces. & \exLink{github.com/ros/executive\_smach}{https://github.com/ros/executive_smach} \\ \hline
FlexBE & Modeling; Execution; Monitoring; Debugging & Behavior engineering toolkit with a graphical \sm\ editor, execution engine, onboard monitoring, and debugging tools. & \exLink{github.com/flexbe}{https://github.com/flexbe} \\ \hline
YASMIN & Modeling; Execution; Monitoring; Debugging & ROS package for designing and executing \sm\ using Python. Provides a shared blackboard for data exchange and a lightweight execution engine. & \exLink{github.com/uleroboticsgroup/yasmin}{https://github.com/uleroboticsgroup/yasmin} \\ \hline
SMACC2 & Execution; Monitoring; Debugging & Event-driven, asynchronous hierarchical state machine library for ROS2 in C++. & \exLink{smacc2.robosoft.ai}{https://smacc2.robosoft.ai/} \\ \hline
RAFCON & Modeling; Execution; Monitoring; Debugging & Graphical tool for hierarchical and concurrent \sm. Includes a graphical interface with visualization, variable inspection, breakpoints, and step-by-step debugging.& \exLink{github.com/DLR-RM/RAFCON}{https://github.com/DLR-RM/RAFCON/} \\
\bottomrule
\end{tabular}
}}
\end{table*}

\begin{obs}{BT-related tools} \label{obs:bt-tools}
The most widely adopted BT tools are the BehaviorTree.CPP ecosystem (including the Groot editor) and the PyTrees suite. These libraries are actively maintained, well-documented, and supported by a large and active community.
However, they adopt a different default execution semantics: BehaviorTree.CPP implements memoryful control nodes, while PyTrees employs a stateless approach. This implementation-level characteristic influences modeling choices and practical usage of the tools.
\end{obs}

{\bf SM}: unlike other formalisms, we did not identify established baseline tools outside the robotics domain that have served as foundations for ROS-compatible solutions (see Table~\ref{tab:tools_sm}).
Instead, the most widely adopted tools in robotics are those developed directly within the ROS ecosystem itself. In particular, \textit{SMACH} and \textit{FlexBE} are mature solutions that provide modeling, execution, and visualizations specifically tailored for robotic behaviors.
Additionally, the \textit{YASMIN} tool~\cite{yasmin-SantamartaLOL22} has been proposed to address the initial lack of ROS2 compatibility in \textit{SMACH} and \textit{FlexBE}. It is actively maintained and provides execution support and modeling facilities for \sm-based robotic mission specification within.
Similarly, \textit{SMACC2} is a ROS2-oriented library designed to address real-world industrial scenarios with real-time requirements. It does not provide graphical modeling support, as state machines are defined directly in code. Nevertheless, it provides execution capabilities, along with monitoring and debugging support, through built-in runtime visualization and diagnostic tools.
Finally, \textit{RAFCON}~\cite{rafconBrunnerSBD16} offers a hierarchical state machine framework that features concurrent state execution for representing complex robot programs. It includes a graphical user interface for creating and editing state machines and provides IDE-like debugging mechanisms to support development and runtime monitoring.

\begin{obs}{SM-related tools} \label{obs:sm-tools}
SM tools have generally been developed independently, without building upon an established baseline tool. Among the most widely adopted are SMACH and FlexBE, while YASMIN and SMACC2 are gaining interest as the community transitions toward the ROS2 framework. Feedback from the evaluation indicated that these tools can present usability challenges, and that SMACH in particular appears to be approaching the end of its life.
\end{obs}

\begin{table*}[!htbp]
\centering
\caption{Identified tools for \htn (Grey-highlighted rows indicate baseline tools supporting the core formalism and used as foundations for ROS-based packages).}
\label{tab:tools_htn}
\resizebox{\linewidth}{!}{
{\footnotesize
\begin{tabular}{p{0.12\linewidth}p{0.13\linewidth}p{0.39\linewidth}p{0.26\linewidth}}
\toprule
\textbf{Name} & \textbf{Scope} & \textbf{Description} & \textbf{Documentation Reference} \\
\midrule \midrule
\rowcolor{mygray} Pyhop & Planning & Lightweight Python \htn\ planner. Suitable for prototyping due to its simple implementation. & \exLink{pyhop}{https://bitbucket.org/dananau/pyhop/src/master/} \\ \hline
\rowcolor{mygray} SHOP family & Planning & Domain-independent automated-planning systems based on ordered task decomposition.  & \exLink{cs.umd.edu/projects/shop}{https://www.cs.umd.edu/projects/shop/description.html} \\  \hline
\rowcolor{mygray} InductorHTN & Planning & Python and C++ lightweight HTN planning engine based on a Prolog compiler. It has been developed and used for providing planning support to agents in mobile games. & \exLink{github.com/EricZinda/InductorHtn}{https://github.com/EricZinda/InductorHtn} \\ \hline
\rowcolor{mygray} HTN Planning AI & Modeling; Planning; Execution; Debugging & Plugin for Unreal Engine providing support for planning game characters' AI. It provides support for modeling HTNs through a graphical interface, generating plans out of it, running, and debugging the plan. & \exLink{maksmaisak.github.io/htn}{https://maksmaisak.github.io/htn} \\ \hline
ROSPlan & Planning; Execution & ROS-integrated planning framework. Generates plans and dispatches them to ROS components. & \exLink{kcl-planning.github.io/ROSPlan}{https://kcl-planning.github.io/ROSPlan/} \\ \hline  
PlanSys2 & Planning; Execution & ROS package providing a planning system. After computing a plan, it automatically converts it into an executable \bt, enabling integration with robot controllers.
ROS planner package that, after obtaining the plan, automatically converts it into an executable \bt. & \exLink{plansys2.github.io}{https://plansys2.github.io/} \\
\bottomrule
\end{tabular}
}}
\end{table*}

\begin{table*}[!htbp]
\centering
\caption{Identified tools for \bpmn (Grey-highlighted rows indicate baseline tools supporting the core formalism and used as foundations for ROS-based packages).}
\label{tab:tools_bpmn}
\resizebox{\linewidth}{!}{
{\footnotesize
\begin{tabular}{p{0.12\linewidth}p{0.13\linewidth}p{0.39\linewidth}p{0.26\linewidth}}
\toprule
\textbf{Name} & \textbf{Scope} & \textbf{Description} & \textbf{Documentation Reference} \\
\midrule \midrule
\rowcolor{mygray} Camunda & Modeling; Execution; Monitoring; Debugging & Process orchestration \bpmn\ platform. Provides graphical modeling tools, workflow engines, REST interfaces, and dashboards for runtime monitoring and debugging. & \exLink{camunda.com}{https://camunda.com/} \\ \hline
\rowcolor{mygray} bpmn.io & Modeling  & Web-based \bpmn\ editor and viewer for designing and visualizing process models. Serves as the front-end basis for many \bpmn-based applications. Provides many add-ons that can extend its scope.
& \exLink{bpmn.io}{https://bpmn.io/} \\ \hline
FaMe & Modeling; Execution; Monitoring & \bpmn-driven framework for multi-robot system development. Provides \bpmn\ modeling, automatic ROS-compliant mission configuration, and an execution engine implemented as a ROS node. & \exLink{pros.unicam.it/fame}{https://github.com/SaraPettinari/fame} \\  \hline
TRACE & Modeling; Execution & A BPMN execution engine providing a connector with the ROS framework. & \exLink{github.com/nasa/trace-executive}{https://github.com/nasa/trace-executive} \exLink{github.com/nasa/trace-ros-connector}{https://github.com/nasa/trace-ros-connector} \\ \hline
B2XKlaim & Modeling; Execution & Translates BPMN diagrams into executable multi-robot coordination code in Klaim, enabling visual mission design and automated code generation. & \exLink{github.com/khalidbourr/B2XKlaim}{https://github.com/khalidbourr/B2XKlaim} \\
\bottomrule
\end{tabular}
}}
\end{table*}

{\bf HTN}: is supported by a set of tools focused on task decomposition and planning (see Table~\ref{tab:tools_htn}). Planners like the \textit{SHOP} family provide mature planning engines that allow the specification of tasks and methods in a hierarchical manner, supporting automated planning and reasoning over complex behaviors.
A few tools, like \textit{InductorHTN} and \textit{Hierarchical Task Network Planning AI} were specifically developed for controlling agents within videogames. The latter, in particular, provides a complete toolset, with a graphical interface, to assist developers in realizing \htns, computing plans (with runtime replanning support), simulating their execution, executing them over the Unreal Engine environment, and debugging.
In robotics, frameworks like \textit{ROSPlan}~\cite{rosplanCashmoreFLMRCPH15} and \textit{PlanSys2}~\cite{plansys2} extend \htn\ planning capabilities to the ROS ecosystem, enabling the integration of task planning within robot behavior. 
We note that ROSPlan natively supports PDDL rather than HTN representations. Nevertheless, HTN models can be translated into PDDL under specific restrictions~\cite{alford2009translating}.
Additionally, only a few prototype repositories can be found, they are tightly coupled to specific experimental setups, such as \exLink{github.com/Robertorocco/Pick\_Place\_Blocksworld\_Environment}{https://github.com/Robertorocco/Pick_Place_Blocksworld_Environment}, or are no longer actively maintained, like \exLink{github.com/Leontes/ros\_htn}{https://github.com/Leontes/ros_htn}, limiting their practical reuse. 
Additionally, a few works, such as~\cite{rodrigues2022architecture,filippone2024handling,gil2023mission}, proposed HTN implementations for execution into ROS-based mission execution. However, they all employ ad hoc realized representations of \htns, representing already-instantiated \htn trees resulting from planning. They are either manually-provided or obtained using one of the aforementioned tools, where all the abstract tasks are already one-to-one bound with methods refining them.

\begin{obs}{HTN-related tools} \label{obs:htn-tools}
Although the SHOP family of planners is relatively dated and no longer actively maintained, many HTN-based tools still build upon it or derive from its principles. In the ROS ecosystem, tools have been primarily developed around PDDL, or aim to integrate planning results with execution mechanisms such as BTs to enable direct deployment in robotic systems.
\end{obs}

{\bf BPMN}: is widely supported by an ecosystem of tools that cover the entire lifecycle of a process model, including process modeling, enactment, and monitoring (see Table~\ref{tab:tools_bpmn}). Several industrial and open-source platforms, such as \textit{Camunda} and \textit{bpmn.io}, provide editors, execution engines, and dashboards that facilitate the design and execution of \bpmn\ workflows. These tools can form the basis for \bpmn\ solutions that specify and execute robotic missions, as they provide infrastructures that can be adapted for robotic applications. However, as this formalism belongs to a business and organizational domain, only a few solutions exist to operationalize \bpmn\ in the robotic domain.
Among them, the \textit{FaMe} framework~\cite{FaMe} provides support for modeling robotic missions and configuring them to be ROS-compliant. It enables mission execution through a \bpmn\ engine implemented as a ROS node, developed by extending the functionalities of the bpmn.io toolsuite.
Similarly, the \textit{TRACE} tool~\cite{de2020event} is intended to support both the modeling and execution of planned and contingent activities in robotic space missions. Unlike most other tools, it also includes model verification capabilities to assess feasibility before execution. However, the publicly available implementation appears to provide only execution functionality and lacks comprehensive documentation.
Finally, \textit{B2XKlaim}~\cite{bourr2026translating} takes a different approach by translating BPMN diagrams into executable multi-robot coordination code in the Klaim language. It enables users to visually design robot missions in BPMN, while the generated Klaim code supports their execution.

We acknowledge the existence of an additional \bpmn-based solution presented \cite{OchoaLLP24}. However, we do not include it among the available tools as it is integrated into a broader and highly specialized workflow-management suite rather than being focused on robotic missions specification and execution, and has an outdated corresponding repository. For these reasons, we mention this work for completeness but do not list it as a usable tool in our analysis.

\begin{obs}{BPMN-based tools} \label{obs:bpmn-tools}
Existing BPMN-ROS solutions build on established tool suites that provide mature execution engines and modeling environments whose functionalities can be extended (e.g., to support ROS-based functionality). 
However, we observed that these tools have mainly been developed within the scope of specific projects or research publications. As a result, their broader adoption and long-term impact on the robotics community require further investigation.
\end{obs}

\subsection{Validation}
To ensure the accuracy and completeness of the tools analysis, we asked the experts to confirm whether they were familiar with or had used the identified tools, to assess the accuracy of their classification in terms of scope and capabilities, and to indicate whether any relevant tools or aspects had been overlooked. Specifically, each expert was asked the following VQs: 
\begin{itemize}[labelwidth=3.2em, labelsep=0.6em, leftmargin=!, align=left]
    \item[VQ3.1] \textit{Are you familiar with or have you previously used any of the listed tools? }
    \item[VQ3.2] \textit{Based on your experience, do you agree with the assigned scope? }
    \item[VQ3.3] \textit{Are there further ROS-related tools that are missing?}
\end{itemize}

The feedback collected during this interview was used to refine and strengthen the results presented in this section and to gather additional insights into the tools presented.
For VQ3.1 and VQ3.3, participants responded to open-ended questions, whereas for VQ3.2, they rated their agreement on a 5-point Likert scale. Specifically, in VQ3.1, participants were asked to indicate which tools from the presented table they were familiar with, or to state ``none'' otherwise. If familiarity with any tool was acknowledged, in VQ3.2, participants evaluated the correctness of the assigned tool scope. Finally, in VQ3.3, participants were invited to suggest additional formalism-related ROS tools that may have been missing from our analysis.
Furthermore, participants were invited to provide feedback on tools they were familiar with, allowing us to capture end-user experiences.

\textbf{Results and discussion}: 
With respect to tool familiarity (VQ3.1), Figure~\ref{fig:tool-mentions-by-formalism} summarizes the distribution of tool knowledge across respondents. To provide a comprehensive view, the figure also includes tools suggested in response to VQ3.3, which are marked with an $\ast$.

For \bt, most respondents are familiar with the \textit{BehaviorTree.CPP} ecosystem, while \textit{PyTrees} is slightly less commonly known. For \sm, familiarity is primarily associated with \textit{SMACH}, although it is known only by approximately half of the respondents (14 out of 27). Regarding \htn, a small subset of respondents reported familiarity with the \textit{SHOP} family, while other solutions appear to be less widely recognized. 
Finally, for \bpmn, the baseline tool suites \textit{Camunda} and \textit{bpmn.io} are the most commonly known among participants.
We also analyzed the proportion of ``none'' responses, i.e., participants who indicated no familiarity with the proposed tools and did not suggest alternatives. This proportion amounts to approximately 27.2\% of \bt experts (6/22), 33.3\% of \sm experts (9/27), 50\% of \htn experts (5/10), and 28.6\% of \bpmn experts (4/14). This distribution suggests that, specifically for \htn, knowledge may be more widespread at a conceptual or theoretical level than at the level of concrete tool usage and adoption.

Concerning the agreement with the assigned tool scope (VQ3.2), respondents expressed overall positive evaluations. Only one participant suggested a modification, noting that YASMIN also provides debugging support. 
This observation has been incorporated into \Cref{tab:tools_bt}, \Cref{tab:tools_sm}, \Cref{tab:tools_htn}, and \Cref{tab:tools_bpmn} and in the corresponding tool description. 
Additionally, another participant raised concerns about ROSPlan’s primary support for PDDL, noting that integrating HTNs would require translation into PDDL. We revised the description of the ROSPlan framework to explicitly clarify this aspect, while keeping its inclusion as a potential solution for integrating HTN-based approaches within the ROS ecosystem.
\Cref{fig:likert-vq3} reports the distribution of agreement levels. The number of respondents considered for this analysis includes only those who indicated familiarity with at least one tool in VQ3.1. Overall, the agreement levels confirm the appropriateness of the adopted scope classification.

Regarding missing tools (VQ3.3), respondents suggested adding additional solutions. 
In particular, \textit{ros2\_ros\_bt\_py} was included among the \bt tools, and \textit{SMACC2} was added to the \sm ones. We also discussed in the text \exLink{github.com/narcispr/py\_trees\_meet\_groot}{https://github.com/narcispr/py_trees_meet_groot}, which provides an automatic translation between Groot and PyTrees, and represents a solution of interest for future investigation.
Furthermore, for \bpmn, \textit{TRACE} and \textit{B2XKlaim} were explicitly mentioned and have now been incorporated into \Cref{tab:tools_bpmn} and discussed in the text. \textit{TRACE} was already known from the literature; however, as its implementation was not initially mapped to a publicly available repository, it was not listed in the original table.

Finally, we report the qualitative feedback collected from respondents regarding their practical experience with the analyzed tools.
With respect to \bts, respondents generally provided positive feedback on the maturity and usability of the \textit{BehaviorTree.CPP} and \textit{PyTrees} ecosystems. However, some participants noted differences in semantics between \textit{BehaviorTree.CPP} and \textit{PyTrees}, suggesting that these differences may influence expressivity and modeling choices. This aspect has been summarized in Observation~\ref{obs:bt-tools}.
Regarding \sms tools, the feedback was more heterogeneous. Several respondents highlighted usability challenges, reporting that some SM tools can be difficult to configure or use in practice. We reported this in Observation~\ref{obs:sm-tools}. At the same time, a contrasting experience was reported (targeting \textit{SMACH} and \textit{FlexBE}), which was described as being in academic environments and relatively easy for students to use. These diverging perspectives suggest that usability may depend significantly on context and usage objectives.
For \htn-related tools, a participant emphasized that the \textit{SHOP} family frameworks, although still considered reference implementations, are relatively old and not actively maintained, as reported in Observation~\ref{obs:htn-tools}.
Finally, for \bpmn, feedback was largely positive. In particular, \textit{bpmn.io} was explicitly appreciated for its flexibility and suitability for adapting BPMN models to mission-specific requirements. This highlights the modeling and execution support provided by \bpmn tools, even if its adoption in robotics remains less widespread.

\begin{observation}
\textbf{Summary of RQ3}.
Publicly available tool support is uneven across the four formalisms. \bts and \sms benefit from several actively maintained, ROS-oriented frameworks that primarily target modeling, execution, monitoring, and debugging. \htn support is more limited and focuses on planners, with only a few tools providing integration with robotic execution. \bpmn is supported by a mature ecosystem of business-process tools with strong modeling, execution, monitoring, and debugging capabilities; however, its integration into robotic systems remains at an early stage.
Feedback from the validation further highlighted differences in usability and maturity: BT tools and BPMN baseline tools were generally perceived as robust and well supported, SM tools were often noted as less user-friendly, and HTN tools were considered comparatively dated and less actively maintained.
\end{observation}

\definecolor{colorBT}{HTML}{3B6EA8}
\definecolor{colorSM}{HTML}{2cb8a7}
\definecolor{colorHTN}{HTML}{f0822e}
\definecolor{colorBPMN}{HTML}{9876c4}

\newcommand{\btcpp}{\shortstack{BehaviorTree.CPP\\BehaviorTree.ROS2}}

\newcommand{\pytrees}{\shortstack{PyTrees\\PyTrees ROS}}

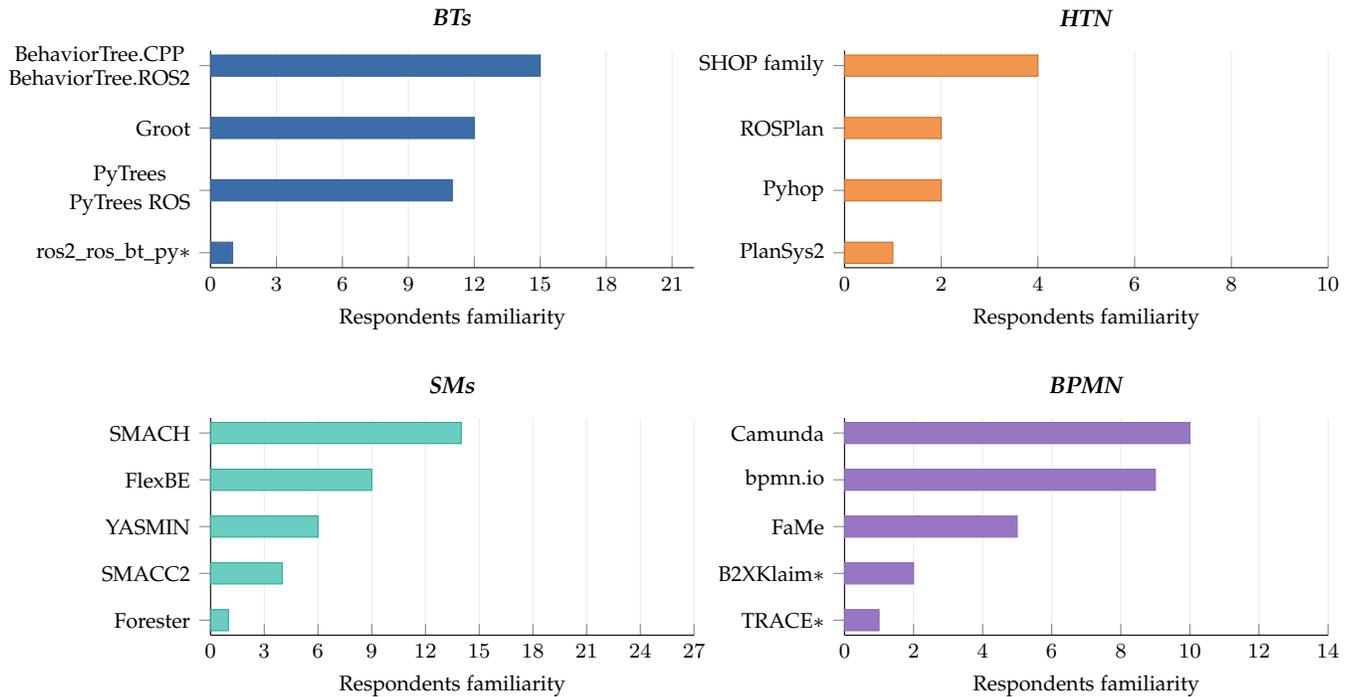
\begin{figure*}[t]
\centering
\begin{tikzpicture}

\begin{groupplot}[
    group style={
        group size=2 by 2,
        horizontal sep=2cm,
        vertical sep=2cm
    },
    xbar,
    width=0.44\textwidth,
    height=0.18\textheight,
    axis x line*=bottom,
    axis y line*=left,
    xmajorgrids,
    grid style={gray!15},
    tick label style={font=\footnotesize},
    label style={font=\footnotesize},
    title style={font=\small\bfseries\itshape, align=center},
    yticklabel style={font=\footnotesize},
    enlarge y limits=0.08,
]

\nextgroupplot[
    title={BTs},
    xmin=0, xmax=22,
    xtick={0,3,6,9,12,15,18,21},
    xlabel={Respondents familiarity},
    symbolic y coords={
        \btcpp,
        Groot,
        \pytrees,
        ros2\_ros\_bt\_py$\ast$
    },
    ytick=data,
    y dir=reverse,
]

\addplot[
    fill=colorBT,
    draw=colorBT!90!black,
    bar width=8pt
] coordinates {
    (15,\btcpp)
    (11,\pytrees)
    (12,Groot)
    (1,ros2\_ros\_bt\_py$\ast$)
};

\nextgroupplot[
    title={HTN},
    xmin=0, xmax=10,
    xtick={0,2,4,6,8,10},
    xlabel={Respondents familiarity},
    symbolic y coords={
        SHOP family,
        ROSPlan,
        Pyhop,
        PlanSys2
    },
    ytick=data,
    y dir=reverse,
]

\addplot[
    fill=colorHTN!85,
    draw=colorHTN!85!black,
    bar width=8pt
] coordinates {
    (4,SHOP family)
    (2,ROSPlan)
    (2,Pyhop)
    (1,PlanSys2)
};

\nextgroupplot[
    title={SMs},
    xmin=0, xmax=27,
    xtick={0,3,6,9,12,15,18,21,24,27},
    xlabel={Respondents familiarity},
    symbolic y coords={
        SMACH,
        FlexBE,
        YASMIN,
        SMACC2,
        Forester
    },
    ytick=data,
    y dir=reverse,
]

\addplot[
    fill=colorSM!70,
    draw=colorSM!90!black,
    bar width=8pt
] coordinates {
    (14,SMACH)
    (9,FlexBE)
    (6,YASMIN)
    (4,SMACC2)
    (1,Forester)
};

\nextgroupplot[
    title={BPMN},
    xmin=0, xmax=14,
    xtick={0,2,4,6,8,10,12,14},
    xlabel={Respondents familiarity},
    symbolic y coords={
        Camunda,
        bpmn.io,
        FaMe,
        B2XKlaim$\ast$,
        TRACE$\ast$
    },
    ytick=data,
    y dir=reverse,
]

\addplot[
    fill=colorBPMN,
    draw=colorBPMN!90!black,
    bar width=8pt
] coordinates {
    (10,Camunda)
    (9,bpmn.io)
    (5,FaMe)
    (2,B2XKlaim$\ast$)
    (1,TRACE$\ast$)
    
};

\end{groupplot}
\end{tikzpicture}

\caption{Tool mentions grouped by formalism ($\ast$-marked tools represent the ones suggested by respondents).}
\label{fig:tool-mentions-by-formalism}
\end{figure*}

\LikertChartVQTwoTwo{20}
{(0,\bt) (0,\sm) (0,\htn) (0,\bpmn)} 
{(0,\bt) (0,\sm) (0,\htn) (0,\bpmn)} 
{(1,\bt) (1,\sm) (1,\htn) (1,\bpmn)} 
{(10,\bt) (12,\sm) (2,\htn) (6,\bpmn)} 
{(5,\bt) (5,\sm) (2,\htn) (3,\bpmn)} 
{Likert responses for VQ3.2}
{fig:likert-vq3}
\section{Threats to Validity}\label{sec:threats}

We identified the main threats to validity and mitigation
strategies according to~\cite{wohlin2012experimentation}, as described in the following.

\textit{External validity.}
Our validation relies on a questionnaire whose respondents were selected through purposive sampling, by recruiting the authors of the reference works underlying the comparison. This choice was motivated by the need to involve experts with direct knowledge of the considered formalisms and their application to robotic systems. However, this may have introduced positive bias, as participants may be more favorable toward the formalisms they helped develop, apply, or popularize.
We mitigated this threat in several ways. First, the main findings of the paper were obtained from the literature, documentation, scenario modeling, and tool analysis, while the questionnaire was only used to validate and refine findings. Second, participants self-assessed their expertise on all four formalisms and answered only the questions related to the formalisms for which they declared sufficient expertise. Thus, participants were not restricted to evaluating only the formalisms associated with their own publications, but could provide feedback on any formalism they knew.
Finally, the questionnaire was not designed to rank or explicitly compare the formalisms. Participants were asked to assess the correctness and completeness of the presented comparison rather than express preferences, thus reducing the risk of preference-driven responses.

\textit{Construct validity.}
A threat to the construct validity of the work concerns the design of the expert questionnaire. Participants were required to provide a textual justification only for neutral or lower Likert scores. This choice was made to make lower-agreement scores interpretable and actionable, and it allowed us to derive qualitative insights to refine the findings and derive observations. However, the additional effort required for neutral or negative answers may have unintentionally encouraged more positive ratings, especially in the presence of questionnaire fatigue.
We mitigated this threat by running a follow-up phase with highly experienced participants to discuss the most relevant findings. 

\textit{Internal validity.}
A further threat concerns the subjectivity involved in interpreting the formalisms, modeling the scenarios, and mapping scenario characteristics to the identified concerns.
We mitigated this threat by involving multiple authors in the modeling and review process (as described in \Cref{sub:methodology-analysis}), cross-checking the produced models, and discussing disagreements internally. Moreover, the questionnaire validation and follow-up interactions with experts were used to further assess the correctness of our findings and refine the comparison where needed.
\section{Discussion}\label{sec:discussion}

This section discusses the implications of our results from two complementary perspectives. First, we reflect on how well the four formalisms capture mission concerns that arise in real-world service and multi-robot settings, including variability introduced by human involvement and runtime uncertainty. Second, we discuss adoption-oriented factors, such as readability, reuse, accuracy, and tooling integration, that often determine whether a formalism is viable beyond controlled examples. Finally, we discuss how software architectures support the operationalization of mission models by mitigating formalism-level limitations, encapsulating integration logic, and enabling mixed-formalism designs.

\subsection{Formalisms expressiveness for real-world needs} 

Missions are a major driver of variability for service robots~\cite{garcia2023software}. In particular, mission specification formalisms should support key sources of variability~\cite{garcia2023software}: (i) the expertise of the human operator varies and is often domain-dependent, (ii) the means of human--robot interaction range from traditional interfaces to gesture- or voice-based interaction, and (iii) humans may share the environment with robots and participate at different levels, from passive to active to proactive involvement~\cite{TOSEM2025}. Human presence, therefore, requires abilities beyond merely completing the mission safely: robots increasingly operate with configurable degrees of autonomy and may need to adapt their behavior to human intentions and, potentially, affective cues (e.g., deciding when control should shift from the robot to the human to avoid ethically problematic situations)~\cite{TOSEM2025}. These aspects pose requirements on mission formalisms: they should be expressive enough, and they should enable operators, given their expertise and available interaction mechanisms, to specify missions correctly, safely, and at an appropriate level of detail~\cite{garcia2023software}.

Within this context, our analysis indicates that the four formalisms exhibit expressiveness gaps that are often addressed through external code or additional artifacts. Besides our results, prior work highlights the advantages of \bts, such as flexibility~\cite{thorsten2023behavior}, reactivity, and modularity~\cite{colledanchise2016advantages}. However, several mission concerns (e.g., interaction logic, temporal constraints, and waiting) are typically delegated to action-node implementations, as the core syntax provides no dedicated constructs for these aspects. This tendency also makes \bts tightly coupled with ``behavioral glue code'' that links the model to the underlying software system~\cite{ghzouli2020behavior}. Moreover, some behavioral aspects (notably concurrency) are not strictly defined by the \bt formalism and are often left to user-defined execution policies and implementations. A similar conclusion applies to \sms, where many concerns are pushed into state implementations and supporting infrastructure.

\htns are less suited to missions where key choices cannot be fully committed at design time because they depend on runtime observations or evolving conditions (e.g., in the FL scenario, some decisions cannot be taken in advance~\cite{filippone2024handling}). While this limitation can be mitigated by providing multiple alternative methods for the same abstract task, this shifts the burden to \emph{online} method selection and (re-)planning, which typically requires dedicated mission-management components (state monitoring, re-planning triggers, and safe plan replacement) and non-trivial engineering effort. The challenge is amplified in multi-robot missions, where decisions depend on distributed state (e.g., teammate availability and communication delays), making consistent replanning and coordinated task allocation harder to implement and validate.

Finally, our findings suggest that the formalisms are better suited to different abstraction levels. \bpmn is well aligned with mission-level modeling, as it supports explicit orchestration and can facilitate the integration of robots with other devices as well as with human workflows. Compared to the other formalisms, the main advantage of \htn is that decision making is specified \emph{declaratively} (tasks, alternative methods, and preconditions), rather than being hard-coded through explicit control-flow; in automated planning terms, this corresponds to \emph{deliberation}, i.e., reasoning about which course of action to take based on goals and the current state~\cite{ghallab2016automated}. The resulting \htn specification can then be executed operationally like a process, similarly to what is done in distributed and architecture-oriented robotics planning frameworks~\cite{lesire2016distributed,rodrigues2022architecture}. In contrast, \bts are often a good fit for task-level specifications, while \sms are typically suitable for well-scoped skills or modes with clear event-driven transitions. Importantly, these formalisms are not mutually exclusive: as also suggested by some interviewees, a pragmatic approach is to combine them so that each is used where it provides the strongest modeling support (e.g., \bpmn for orchestration and human/robot handoffs, \htn for deliberation and plan synthesis, and \bts/\sms for reactive execution and skill-level control).

\subsection{Formalisms adoption in real-world settings}\label{sec:adoption}

Beyond expressivity (\Cref{sec:concepts}), additional factors influence whether a mission specification formalism is adopted in practice, especially when intended users may include domain experts rather than robotics specialists. In what follows, we discuss four recurrent adoption drivers: \emph{(i) simplicity} (including complexity and readability), \emph{(ii) scalability}, \emph{(iii) extensibility}, \emph{(iv) reusability}, \emph{(v) accuracy}, and \emph{(vi) integrability} with available tooling and with heterogeneous devices (e.g., IoT sensors).

{\em Simplicity, complexity, and readability}. 
We interpret \emph{complexity} as the number of modeling elements and the amount of connections, relationships, and interdependencies that must be managed to specify a mission, and \emph{readability} as the extent to which a human can understand, maintain, and debug the model without extensive effort, both aspects being central for adoption~\cite{IovinoFFCSS25}. Among the considered formalisms, \bts are often perceived as less immediate for non-experts: their tick-based semantics make the control flow implicit and not directly readable from the tree shape, so understanding the runtime behavior often requires mentally simulating the ticking mechanism and the propagation of status values. In addition, some constructs can become fragile when striving for robustness: since both conditions and actions return the same status domain, failures can trigger the same fallback behavior, which may inadvertently steer execution to alternative branches unless additional guarding logic is introduced. As a result, seemingly simple concerns (e.g., implementing a \emph{wait} or carefully isolating failure causes) may lead to non-trivial modeling patterns and increased structural complexity. \htns, while also tree-shaped, often offer a more linear ``reading'' of execution (e.g., depth-first decomposition), which can make the intended flow easier to grasp at a high level. In contrast, \sms and \bpmn expose control flow more explicitly through their transition-/token-based semantics, which often improves traceability of execution paths. At the same time, \bpmn provides a rich and standardized notation, which can support communication and documentation, but its breadth of constructs can raise the entry barrier and require deeper familiarity; this suggests that \bpmn may be particularly suitable for documenting and communicating mission workflows at higher abstraction levels.

{\em Scalability}. 
In terms of scalability, i.e., the ability to model increasingly large and detailed missions while preserving readability and manageability, hierarchical structuring and modularization are key. \bts typically scale well through subtree composition and reuse, although scalability degrades when many mission concerns are implemented inside action nodes, splitting the logic between the model and code. SMs can become difficult to maintain as missions grow due to state/transition explosion, even if hierarchy mitigates it partially. HTNs scale effectively as domain libraries by adding methods/operators, but large method sets increase maintenance and debugging effort and often require additional infrastructure (e.g., monitoring and replanning) to remain reactive. BPMN can remain readable at scale when using subprocess decomposition and clear conventions, but highly detailed robot behaviors may still lead to large process models and extra effort to connect process-level logic to execution-level mechanisms.

{\em Extensibility}. 
Extensibility is a desired characteristic of mission specification formalisms~\cite{IovinoFFCSS25}: as missions evolve, engineers need to introduce new behaviors and cross-cutting concerns without rewriting large portions of the model. \bts can often support localized extensions by composing or replacing subtrees and by using decorators to wrap existing behaviors (e.g., retries, timeouts, guards, and recovery) with limited impact on otherwise independent parts. In contrast, some formalisms may ``explode'' under incremental change: \sms can suffer from state/transition explosion when new interrupts, priorities, and exception paths must be integrated across many states, while BPMN models can become unwieldy if low-level contingencies and exception handling are explicitly encoded in the process. This motivates combining complementary models so that extensions remain localized to the abstraction layer they affect.

Multi-robot settings further stress extensibility because they introduce task interruption, save-and-resume behavior, and team-level coordination concerns. For example, when priorities change due to limited resources (e.g., battery level), robots may need to interrupt one task and later resume it from the previous computational state; none of the analyzed formalisms support task save-and-resume as a native capability, making resumption logic an additional engineering concern. Moreover, multi-robot missions inherently require \emph{task assignment} and \emph{coordination}: tasks must be allocated based on capabilities and availability, and execution must be synchronized through dependencies, rendezvous points, mutual exclusion over shared resources, and communication protocols. These concerns are rarely captured end-to-end by a single formalism, and extending missions in practice often entails evolving not only the behavior model but also the surrounding coordination mechanisms that ensure coherent team-level execution.

{\em Reusability}.
We define \emph{reusability} as the extent to which mission fragments can be reused across missions with minimal adaptation. In principle, all formalisms support some form of modularization (e.g., hierarchical states, sub-processes, subtree composition, method libraries), but practical reuse depends on how well models can be separated from system-specific code and tooling. For \bts, reuse is often advertised through subtree composition and libraries of nodes; however, empirical evidence indicates that reuse mechanisms in robotics projects are frequently simple and that models may be deeply intertwined with ``behavioral glue code'' connecting them to the underlying software system, which hinders reuse and makes model-level manipulation (visualization, testing, reuse outside the original system) more difficult~\cite{ghzouli2020behavior}. Similar risks exist for other formalisms whenever mission logic is split between the model and extensive external code (e.g., action implementations, event dispatchers, or custom runtime adapters), reducing portability of reusable fragments.
As a standardized notation, \bpmn\ provides explicit constructs for modularization, such as sub-processes and call activities, which can facilitate reuse when models are designed accordingly~\cite{FaMe}. Nevertheless, the actual degree of reuse depends on modeling discipline: tight coupling with specific execution engines or custom extensions may limit portability, whereas well-structured and implementation-agnostic models can be more easily reused across missions and systems.

{\em Accuracy}. 
We interpret \emph{accuracy} as the ability of the formalism to capture relevant mission concerns without relying on undocumented assumptions or external artifacts that are essential for understanding the intended behavior. In real deployments, accuracy is challenged whenever crucial concerns are systematically delegated to low-level implementations (e.g., synchronization protocols, timing/waiting policies, interaction contracts), because the model ceases to be a self-contained representation of the mission. This is particularly problematic for validation and assurance: stakeholders may read the model as complete, while key behaviors are implicitly defined elsewhere. From this point of view, all the considered formalisms are exempt from accuracy limitations, since all of them, in different ways, delegate the expression of concerns to ad hoc behavior implementation or to the underlying execution infrastructure. For instance, \bts and \sms are often used in the implementation of action nodes or state behaviors to encode coordination and communication mechanisms or timing constraints. \htn externalize reactive behavior to the planner. \bpmn, while explicitly modeling many concerns, typically assumes engine-level mechanisms for task suspension and resumption.

{\em Integrability}.
Finally, \emph{integrability} concerns the ease of embedding the specification into an operational robotic system and its ecosystem, leveraging existing tools and interfaces without requiring extensive bespoke adapters. This includes runtime execution support, monitoring/debugging facilities, and interoperability with external devices and services. \bts often integrate high-level decision-making with low-level control through mature robotics-oriented libraries and runtime infrastructures. \bpmn is well supported in business-process platforms and can naturally integrate with enterprise and IoT ecosystems; however, a known limitation is that business processes are typically specified a priori and can behave like rigid action plans at runtime~\cite{flexibilityBook}. Recent proposals, therefore, combine process execution with automated planning to recover from exceptional situations and preserve progress during execution~\cite{MalburgHB23,MarrellaMS17,Marrella19}. In particular, automated planning (including \htn-based approaches) has been argued to be well suited to synthesize at runtime the content of underspecified activities, i.e., generating sub-processes of appropriate granularity when it becomes clear what must be done at a given point in the process~\cite{Marrella19}.

\subsection{Architecture-based strategies for formalism integration}

The results of RQ2 show that several mission concerns are not uniformly supported by the considered formalisms, as also remarked by Observation~\ref{obs:implementation}. In particular, multi-robot interactions, human-robot interactions, and state saving and task resuming often require implementation-level mechanisms for their realization. As remarked in \Cref{sec:expressivity}, a lack of native support in a formalism does not imply that the corresponding concern is not supported by the system, while concerns may still be addressed by the surrounding software architecture and skills implementations.

Indeed, once a mission model is defined, additional architectural components are typically needed, for instance, to connect the model to robot-specific capabilities, allocate tasks to robots, coordinate task execution, and manage communication. Thus, even if tools for executing mission models exist, they must still be embedded within a broader robotic software architecture.
A commonly adopted strategy is to separate mission specification from execution infrastructure. Architecture-based strategies distinguish high-level mission logic from platform-specific control, task allocation, skill invocations, coordination, communication, and adaptation~\cite{andreasson2023softwarearchitectures,ahmad2016roboticarchitectures,rodrigues2022architecture,filippone2024handling,seiger2026autonomous,braberman2015morph,zudaire2022assured}. In such architectures, the mission model does not need to encode every coordination and adaptation mechanism explicitly, since it can rely on external components such as planners, dispatchers, execution engines, and middleware adapters. This is particularly relevant in multi-robot systems, where task allocation and coordination may be managed by dedicated mission-management components independently of the adopted formalism~\cite{rodrigues2022architecture}.

Architectural solutions can be leveraged to mitigate the glue code problem. Instead of embedding robot-specific skill code in mission models, integration logic can be encapsulated in adapters, wrappers, or connectors. For example, BT action nodes, SM states, HTN primitive tasks, and BPMN service tasks can invoke robot-specific functionalities (e.g., ROS actions) or external services through well-defined interfaces exposed by connectors or adapters, hence abstracting the low-level skill implementation.
The same approach can be leveraged to provide reusable implementations for the functionalities that are not directly supported by the formalism (e.g., communication, task resuming, waiting), as discussed in \Cref{sec:expressivity}. In fact, execution-layer components or external services can provide the missing functionality, while the mission model remains focused on the high-level specification and execution.

Architectural-based solutions can also enable the integration of multiple formalisms. For instance, BPMN can be used for high-level mission orchestration and human/external-system interaction, while HTNs can be employed for deliberative task decomposition and allocation. Instead, BTs or SMs can be used to model low-level, reactive task execution.
However, such a combination requires explicit mappings between models, execution layers, and robot capabilities. Thus, part of the complexity may shift from the mission model to the integration infrastructure, where, e.g., a BPMN task can be realized by invoking the execution of a BT or an SM, rather than connecting the task directly to the implementation of a robotic skill. This solution would also foster separation of concerns between the specification of higher-level mission goals and the lower-level reactive behavior of the system.

\section{Related Works}\label{sec:related}

Among the considered formalisms, \bts have attracted the most significant attention in robotics as a modular and reactive formalism for structuring robot behaviors. Prior work has discussed \bts from both conceptual and practical perspectives, including their modeling principles, typical control-flow constructs, and the engineering motivations behind their adoption in robotic systems. In particular, Ghzouli et al.~\cite{ghzouli2020behavior} analyze key \bt characteristics and modeling concepts, relate them to UML state and activity diagrams through a language-level mapping, and complement this discussion with an empirical analysis of how \bts are used in practice by mining GitHub repositories (e.g., adopted libraries, language elements, and reuse patterns). Complementary works provide broader background on BTs in robotics and their benefits: Colledanchise and \"Ogren discuss advantages of \bts over alternative control architectures~\cite{colledanchise2016advantages} and further elaborate on \bt design principles and expressiveness considerations~\cite{colledanchise2021on}, while Iovino et al.~\cite{iovino2022survey} offer a survey of \bts in robotics that synthesizes common patterns of use, implementation practices, and recurring challenges.

Beyond \bt-focused studies, several contributions explicitly compare \bts with \sms, highlighting differences in execution semantics and their practical implications. Berger et al.~\cite{thorsten2023behavior} compare \bts and \sms through the lens of widely used DSL-based implementations (e.g., BehaviorTree.CPP and PyTrees for \bts, and SMACH and FlexBE for \sms), contrasting their modeling constructs and semantics and further analyzing their adoption in open-source projects mined from GitHub. In addition to conceptual and tooling-oriented comparisons, controlled empirical evidence has been reported on the effects of using \bts versus \sms in robot mission specification tasks~\cite{dragule2025effects}, offering a user-centric view on the trade-offs between the two formalisms. Finally, comparative discussions have also been extended to broader mission-specification perspectives that position \bts and \sms with respect to complementary modeling approaches and their support for mission concerns~\cite{IovinoFFCSS25}.

In summary, existing work establishes \bts as a practical and widely adopted formalism in robotics~\cite{ghzouli2020behavior,iovino2022survey,colledanchise2016advantages,colledanchise2021on} and clarifies key trade-offs between \bts and \sms~\cite{thorsten2023behavior,dragule2025effects,IovinoFFCSS25}. Our work builds on these foundations by adopting a mission-specification viewpoint and extends the comparative analysis beyond \bts and \sms to cover additional formalisms and the mission concerns they (explicitly or implicitly) support.
\section{Conclusions and Future Work}\label{sec:conclusion}

This paper compares four mission specification formalisms for robotics, namely \bts, \sms, \htns, and \bpmn, to clarify their expressiveness for real-world missions and the implications for adoption. We address three research questions by analyzing how each formalism represents core control structures and mission concepts across representative scenarios, synthesizing their strengths and weaknesses with respect to recurring mission concerns, and validating the analysis through an expert questionnaire survey complemented by targeted follow-up interactions.

Our results show that the formalisms are complementary rather than interchangeable: each is strongest at a particular abstraction level, while other concerns are offloaded to external artifacts or implementation code. \bpmn best supports mission-level orchestration and integration with human workflows and heterogeneous devices; \htn supports declarative decision making (deliberation) and can synthesize executable structures from tasks, methods, and preconditions; \bts suit task-level reactive execution and modular composition; and \sms fit well-scoped skills and mode-based control. However, key mission concerns (e.g., temporal constraints, waiting, interaction protocols, and aspects of concurrency) are frequently delegated to action/state implementations, reducing the model’s self-containment. In multi-robot missions, task interruption with save-and-resume, task assignment, and coordination over distributed state remain largely unsupported as first-class constructs and typically require additional mission-management infrastructure.

A key takeaway is therefore that these formalisms should be viewed as \emph{complementary rather than competing}. In practice, combining them can be a pragmatic strategy to keep specifications readable, maintainable, and evolvable: for example, using \bpmn or \htn at the mission level for orchestration and deliberation, while delegating execution-level robustness to \bts and skill/mode logic to \sms. This layered use also helps mitigate scalability and extensibility issues that may arise when a single formalism is stretched across all mission concerns.

Future work should focus on (i) principled guidelines for multi-formalism mission specifications and their interfaces (e.g., plan-to-execution dispatch, monitoring feedback, and recovery), (ii) reusable patterns and tool support for recurring concerns such as interruption/resumption, distributed coordination, and human-in-the-loop adaptation, and (iii) shared benchmarks and empirical studies on larger systems to quantify trade-offs in scalability, maintainability, and correctness across formalisms and combinations thereof. Moreover, further work should also investigate the computational and memory scalability of the considered formalisms and tools, especially in resource-constrained robotic systems.
This will require implementing and running comparable missions on multiple tools implementing the same formalism, as well as tools across different formalisms, and measuring runtime overhead, memory usage, communication costs, and responsiveness to events or replanning.

\section*{Acknowledgments}
This work has been partially funded by 
(a) the MUR (Italy) Department of Excellence 2023 - 2027, 
(b) the European HORIZON-KDT-JU research project MATISSE ``Model-based engineering of Digital Twins for early verification and validation of Industrial Systems", HORIZON-KDT-JU-2023-2-RIA, Proposal number:  101140216-2, KDT232RIA\_00017, and
(c) the Space It Up project funded by the Italian Space Agency and the Ministry of University and Research CUP: D53C24000580005.

\bibliographystyle{IEEEtran}
\bibliography{references}

\newpage

\section*{Biography Section}
\begin{IEEEbiography}[{\includegraphics[width=1in,height=1.25in,clip,keepaspectratio]{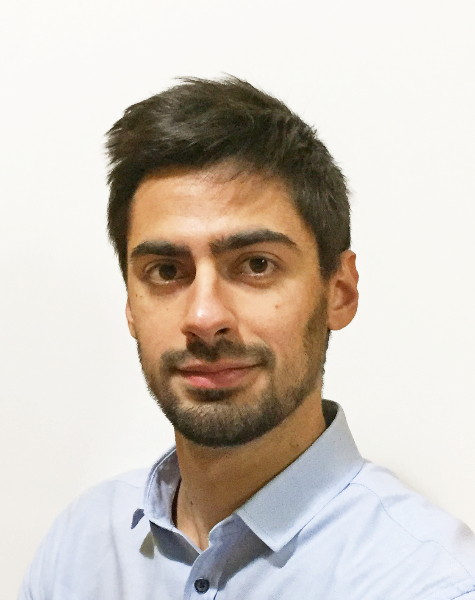}}]{Gianluca Filippone}
is a Postdoctoral researcher in Computer Science at Gran Sasso Science Institute (GSSI, Italy). He received his PhD in Computer Science from the University of L'Aquila, Italy, in 2023. His research topic is software engineering, with focus on autonomous, self-adaptive, and robotic systems. His work spans from service-oriented and distributed architectures for self-adaptive systems to software engineering approaches for the specification and adaptation of robotic and multi-robot missions.
\end{IEEEbiography}

\begin{IEEEbiography}[{\includegraphics[width=1in,height=1.25in,clip,keepaspectratio]{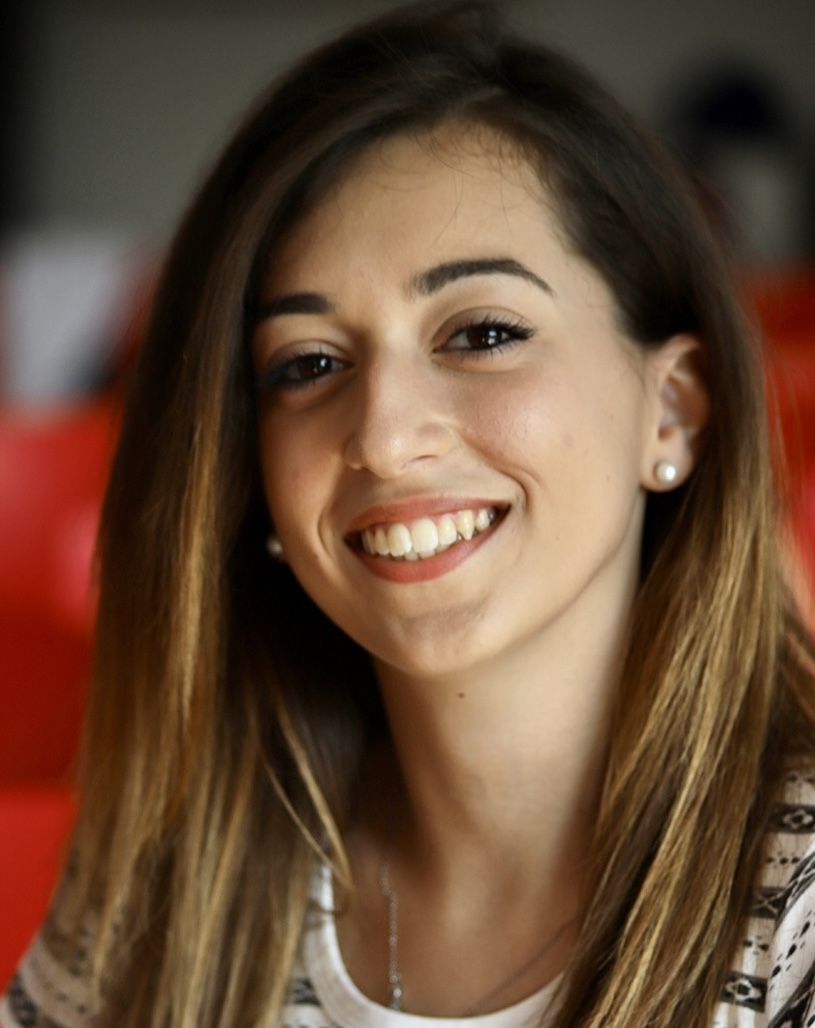}}]{Sara Pettinari}
is a Postdoctoral Researcher in Computer Science at the Gran Sasso Science Institute (GSSI, Italy). She earned her PhD in Computer Science from the University of Camerino. Her research focuses on business process management and process mining, particularly for developing and analyzing robotic systems. Additionally, her work explores the integration of ethical aspects in the design and development of autonomous systems.
\end{IEEEbiography}

\begin{IEEEbiography}[{\includegraphics[width=1in,height=1.25in,clip,keepaspectratio]{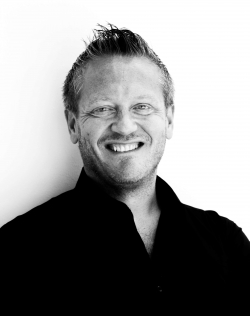}}]{Patrizio Pelliccione}
is a Professor in Computer Science at Gran Sasso Science Institute (GSSI, Italy) and Director of the Computer Science area. Patrizio is also adjunct professor at the University of Bergen, Norway. His research topics are mainly in software engineering, software architecture modeling and verification, autonomous systems, and formal methods. He received his PhD in computer science from the University of L'Aquila (Italy). Thereafter, he worked as a senior researcher at the University of Luxembourg in Luxembourg, then assistant professor at the University of L'Aquila in Italy, then Associate Professor at both Chalmers $\vert$ University of Gothenburg in Sweden and University of L'Aquila.
He has been on the organization and program committees for several top conferences and he is a reviewer for top journals in the software engineering domain. He is very active in European and National projects. In his research activity, he has collaborated with several companies. More information is available at http://patriziopelliccione.com.
\end{IEEEbiography}

\vfill

\end{document}